\documentclass[12pt,a4paper,oneside]{article}

\usepackage[T1]{fontenc}

\usepackage{lmodern}

\usepackage[official,right]{eurosym}
\usepackage{textcomp}   
\usepackage{gensymb}
\usepackage[utf8]{inputenc}
\usepackage[english]{babel}

\usepackage{MnSymbol}
\usepackage[tight,nice]{units}
\usepackage{xspace}
\usepackage{pbox}
\usepackage{siunitx}

\usepackage{titlesec}
\usepackage{afterpage}
\usepackage[pdftex]{graphicx,epsfig}

\usepackage{hyperref}

\usepackage{titlesec}

\begin{document}

\begin{center}
{\large \bf  
Improved search for neutron to mirror-neutron oscillations
in the presence of mirror magnetic fields 
with a dedicated apparatus at the PSI UCN source.}
\end{center}

{\begin{center}
N.J.~Ayres$^{1,*}$, 
Z.~Berezhiani$^{6,7}$,
R.~Biondi$^{6}$,
G.~Bison$^{2}$,
K.~Bodek$^{4}$,
V.~Bondar$^{1}$,
P.-J.~Chiu$^{2,\dagger}$, 
M.~Daum$^{2}$,
R.T.~Dinani$^{5}$,
C.B.~Doorenbos$^{2,\dagger}$,
S.~Emmenegger$^{1}$,
K.~Kirch$^{1,2}$,
V.~Kletzl$^{2,\dagger}$,
J.~Krempel$^{1}$,
B.~Lauss$^{2,*}$
D.~Pais$^{2,\dagger}$,
I.~Rien\"acker$^{2,\dagger,*}$,
D.~Ries$^{2}$,
N.~Rossi$^{6}$,
D.~Rozpedzik$^{4}$,
P.~Schmidt-Wellenburg$^{2}$,
K.S.~Tanaka$^{2}$,
J.~Zejma$^{4}$,
N.~Ziehl$^{1}$,
G.~Zsigmond$^{2}$
\end{center}



\begin{center}
$^{1}$ ETH Z\"urich, Switzerland\\
$^{2}$ Paul Scherrer Institut, Villigen, Switzerland\\
$^{3}$ Department of Chemistry – TRIGA site, Johannes Gutenberg University Mainz, Mainz, Germany\\
$^{4}$ Jagiellonian University, Cracow, Poland\\
$^{5}$ Katholieke Universiteit, Leuven, Belgium\\
$^{6}$ INFN, Laboratori Nazionali del Gran Sasso, Assergi AQ, Italy\\
$^{7}$ Dipartimento di Fisica e Chimica, Universit\'a  di L'Aquila, L'Aquila, Italy\\
$^\dagger${also at ETH Z\"urich}\\
$^*$ Corresponding authors:
Nicholas Ayres, ayresn@phys.ethz.ch,
Bernhard Lauss,
bernhard.lauss@psi.ch,
and
Ingo Rien\"acker,
ingo.rienaecker@psi.ch\\
\end{center}






\vspace*{10mm}
Abstract:\\
While the international nEDM collaboration at the Paul Scherrer Institut (PSI)
took data in 2017 that 
covered a considerable fraction of the parameter 
space of claimed potential signals 
of hypothetical neutron ($n$) to mirror-neutron ($n'$) transitions, 
it could not test all claimed signal regions 
at various mirror magnetic fields.
Therefore, 
a new study of $n-n'$ 
oscillations using stored ultracold neutrons (UCNs)
is underway at PSI,
considerably expanding the reach in parameter space of 
mirror magnetic fields ($B'$) and 
oscillation time constants ($\tau_{nn'}$). 
The new apparatus is designed to test for the anomalous loss 
of stored ultracold neutrons as a function of an applied magnetic field. 
The experiment is distinguished from its predecessors by its 
very large storage vessel (1.47\,m$^3$), enhancing its statistical sensitivity.
In a test experiment in 2020 we have demonstrated the capabilities
of our apparatus. However, the full analysis of our 
recent data is still pending.
Based on already demonstrated performance, 
we will reach a sensitivity to oscillation 
times $\tau_{nn'}/\sqrt{\cos(\beta)}$ well above hundred seconds, with
$\beta$ being the angle between $B'$ and the applied magnetic field $B$. 
The scan of $B$ will allow 
the finding or 
the comprehensive exclusion of potential 
signals reported in the analysis of previous experiments and 
suggested to be consistent with neutron to mirror-neutron oscillations.






\section{Motivation}
\label{Sec:Motivation}



%
%
Already in the seminal publication by Lee and Yang on the 
possibility of the violation of parity symmetry to explain 
some observed phenomena~\cite{Lee:1956qn}, 
the authors pointed out the possibility of an additional symmetry, 
today know as mirror symmetry, 
which motivates the existence of mirror particles.

Parity can be interpreted as a discrete symmetry which exchanges ordinary left-handed particles with their right-handed mirror partners.
Kobzarev, Okun and Pomeranchuk~\cite{Kobzarev:1966qya} 
observed that mirror particles could not have 
ordinary strong, weak or electromagnetic interactions, 
and so they would have to form a hidden 
parallel sector as an exact duplicate of ordinary 
matter, with which it interacts only via gravity.
Thus, all known particles: the electron $e$, proton $p$, neutron $n$, neutrinos $\nu$ etc.\ 
could have mirror partners: $e'$, $p'$, $n'$, $\nu'$ with exactly the same masses, 
being sterile to known strong and electroweak interactions, but having their own 
interactions with each other, in the same ways as their Standard Model partners. 
This idea was applied as an extension to the Standard Model 
in~\cite{Foot:1991bp}  
and worked on by various authors over the decades 
(for a review, see e.g.~\cite{Berezhiani:2003xm,Berezhiani:2005ek,Foot:2014mia}; 
for a historical overview, see also~\cite{Okun:2006eb}).

These thoughts have received renewed interest over the past 20 years, 
as it was shown that  mirror matter could provide 
viable candidates for the dark matter in 
the Universe~\cite{Berezhiani:2000gw,Ignatiev:2003js,
Berezhiani:2003wj}
(more extended discussions of its properties can be found 
in reviews~\cite{Berezhiani:2003xm,Berezhiani:2005ek,Foot:2014mia}).
The interest further increased because of the absence of other clear signals 
of dark matter particles and the ability to explain a number of difficulties in 
collisionless cold dark matter models.
Since mirror matter is atomic/dissipative 
dark matter it can form dark stars and other compact 
objects~\cite{Berezhiani:1995am,Berezhiani:1996sz,Berezhiani:2000gw,Berezhiani:2005vv}.

Plausible cross-interactions between mirror and ordinary
particles, induced e.g. 
by the photon mirror-photon kinetic mixing~\cite{Holdom:1985ag,Glashow:1985ud} 
would, in spite of severe cosmological~\cite{Berezhiani:2008gi}  
and experimental~\cite{Vigo:2019bou} limits,
provide an efficient mechanism for inducing a galactic 
magnetic field~\cite{Berezhiani:2013dea}.
They 
could be revealed via direct detection in dark matter 
detectors~\cite{Foot:2008nw,Cerulli:2017jzz,Addazi:2015cua}.  
In addition, ordinary and mirror sectors can have common gauge interactions related to 
flavor symmetry \cite{Berezhiani:1996ii,Belfatto:2018cfo,Belfatto:2019swo} and 
$B-L$ symmetry \cite{Addazi:2016rgo,Babu:2016rwa}, or common Peccei-Quinn symmetry 
\cite{Rubakov:1997vp,Berezhiani:2000gh}.
A distinct feature  of this concept is that some of these cross-interactions 
may violate baryon (and/or lepton) numbers of both ordinary and mirror sectors. 
At first, these interactions, satisfying Sakharov's conditions~\cite{Sakharov:1967dj}, 
can give rise to co-baryogenesis (or co-leptogenesis) mechanisms in the early Universe,  
which can be the origin of baryon asymmetries in both sectors~\cite{Bento:2001rc,Bento:2002sj}. 
They also could explain the observed cosmological fractions of visible and dark matter components 
in the Universe~\cite{Berezhiani:2008zza,Berezhiani:2018zvs}. 
Secondly, these interactions can induce oscillation phenomena between 
ordinary and mirror particles. In fact, any neutral particle, elementary 
(e.g. neutrinos) or 
composite (e.g. neutrons), can have a mixing with its mass degenerate mirror partner and 
thus can oscillate into the latter. 
In particular, ``active-sterile'' mixings between three 
ordinary neutrinos $\nu_{e\mu\tau}$ and three mirror neutrinos  $\nu'_{e\mu\tau}$
were discussed in Refs.~\cite{Akhmedov:1992hh,Foot:1995pa,Berezhiani:1995yi}.

The possibility of the neutron mass mixing with the mirror neutron
was proposed in Ref.~\cite{Berezhiani:2005hv}.  The phenomenon of $n-n'$ oscillation 
is similar to the neutron--antineutron oscillation $n-\bar n$ 
(see Ref.~\cite{Phillips:2014fgb}
for a review). 
Both phenomena of $n-n'$ and $n-\bar n$ mixing might originate from the 
same mechanism in the context of theoretical models~\cite{Berezhiani:2005hv}. 
However, there are many substantial differences: 
\begin{enumerate}
\item $n-\bar n$ mixing violates baryon number by two units ($\Delta {\rm B}=2$) 
whereas $n-n'$ mixing violates it by one unit ($\Delta {\rm B}=1$).
\item The mass degeneracy between the neutron and antineutron 
stems from CPT invariance. Between the neutron and mirror neutron it 
is a consequence of mirror parity which in principle can be spontaneously broken.
\item Existing limits on the characteristic
$n-\bar n$ oscillation time are rather stringent. 
Namely, the 
experimental direct limit on the $n-\bar n$ oscillation time is 
$\tau_{n\bar n} >0.86 \times 10^{8}$\,s~\cite{Baldo1994}.
Indirect limits from nuclear stability are even stronger,  
$\tau_{n\bar n} >2.7 \times 10^{8}$\,s~\cite{Abe2011}.
As for $n-n'$ oscillation, its characteristic time $\tau_{nn'}$ 
can be as low as a few seconds, 
and in any case much smaller than the neutron lifetime, without 
contradicting   
either existing astrophysical and cosmological limits or,  
unlike  $n-\bar n$ oscillations, nuclear 
stability limits~\cite{Berezhiani:2005hv}. 
The reason why such fast $n-n'$ oscillations are not directly 
manifested experimentally in the neutron losses, 
is that in normal conditions it is suppressed by environmental factors 
as the presence of matter and/or 
magnetic fields~\cite{Berezhiani:2005hv,Berezhiani:2008bc}.
\end{enumerate}

Moreover, such fast $n-n'$ oscillations can have interesting and testable implications 
for ultra-high energy cosmic rays~\cite{Berezhiani:2006je,Berezhiani:2011da}, 
for the neutrons from solar flares \cite{Mohapatra:2005ng},
for neutron stars~\cite{Berezhiani:2020zck,Goldman:2019dbq,Ciancarella:2020msu}, 
for primordial nucleosynthesis~\cite{Coc:2014gia}, 
and can also potentially explain 
the neutron lifetime anomaly~\cite{Greene2016,Berezhiani:2018eds}. 
In addition, it requires the existence of 
BSM physics off the beaten track, e.g.\ color scalar particles, 
at the scale of a few TeV, which can be tested at the LHC and future 
accelerators~\cite{Berezhiani:2005hv,Berezhiani:2015afa}. 
In a more general picture, the neutron $n$ can have mixings with 
both the mirror neutron 
$n'$ and mirror antineutron $\bar{n}'$, with respective masses 
$\epsilon_{nn'}$ and 
$\epsilon_{n\bar n'}$ 
which in combination can induce
neutron--antineutron oscillation at the second order via the
oscillation chain 
$n\rightarrow n', \bar{n}' \rightarrow \bar{n}$, with a characteristic timescale 
$(\tau_{nn'} \tau_{n\bar n'})^{1/2}$ which can be much shorter than the direct 
$n-\bar n$ oscillation time $\tau_{n\bar n}$  \cite{Berezhiani:2020vbe}. 

In 2006 it was pointed out that  
direct experimental limits on $n-n'$ oscillations did virtually 
not exist~\cite{Berezhiani:2005hv}, 
indicating the possibility of an oscillation time $\tau_{nn'}$ 
as short as 1\,s. 
It was suggested to search for these oscillations in laboratories 
via the neutron disappearance $n\rightarrow n'$ or regeneration  $n\rightarrow n' \rightarrow n$
(see also~\cite{Pokotilovski:2006gq,Berezhiani:2017azg}).

\subsection{Previous experimental efforts}

Since then, several experimental collaborations embarked on 
measurements in search of $n-n'$ oscillations, mostly looking for signals of 
neutron disappearance in storage experiments with ultracold neutrons (UCNs). 
Most prominently, two collaborations equipped with setups to search 
for the neutron electric dipole moment provided the relevant limits. 
%
The first experimental limit $\tau_{nn'} > 103$\,s ($95~\%$ C.L.)~\cite{Ban:2007tp} was obtained at ILL 
by the nEDM collaboration that is today working at PSI
and has considerable overlap with the authors of the here described work. 
The strongest experimental limit $\tau_{nn'} > 414$\,s ($95~\%$ C.L.) was obtained at ILL 
by the collaboration centered around the PNPI team of A.~Serebrov
~\cite{Serebrov:2007gw}.
These results were obtained by assuming, as in the original paper~\cite{Berezhiani:2005hv}, 
that there are no effects of mirror matter at the Earth which could affect $n-n'$ oscillations. 
Therefore, these experiments compared the UCN losses in conditions of vanishing ($B\ll 10$~nT) 
and large enough ($B> 1$~$\micro$T) magnetic fields.
These thus  
assumed that $n$ and $n'$ would be degenerate 
around $B=0$ while the application of a magnetic field $B$ of a few $\micro$T would already 
lift the degeneracy and effectively suppress potential oscillations.

Mirror magnetic fields can be induced
by a rather tiny amount of mirror matter~\cite{Ignatiev:2003js}
which can be captured by the Earth,
due to the mirror electron drag mechanism which the Earth rotation may cause.
The mirror magnetic field can reach values comparable to the normal magnetic 
field of the Earth, up to hundreds of $\micro$T. 
In the case  of non-zero $B'$,  
$n-n'$ oscillation at $B\approx 0$  will be suppressed, 
but it would be resonantly enhanced 
if $B \approx B'$. 
Therefore, for finding $n-n'$ oscillation effects, the value of the applied magnetic 
field should closely match the unknown value of $B'$.

Unlike the normal magnetic 
field, the mirror field cannot be screened in experiments, and its direction is also 
unknown. As it was suggested in~\cite{Berezhiani:2008bc}, these effects can be searched for
by measuring UCN losses at different values of the applied field {\bf $B$}.
Then, one can measure the directional asymmetry 
$A_{\uparrow\downarrow}$
by comparing the UCN counts 
between cycles with opposite directions of the applied field at the given value 
($+B$ and $-B$). 
\begin{equation}
   A_{\uparrow\downarrow} = \frac{n_\uparrow - n_\downarrow}{n_\uparrow + n_\downarrow},
	\label{eq:asymmetry}
\end{equation}

where $n_{\uparrow,\downarrow}$ represents the UCN counts $N_{\uparrow,\downarrow}$ 
after a UCN storage time $t_s$,
normalized with the corresponding monitor counts $M_{\uparrow,\downarrow}$
(see Sec.~\ref{subsec:Concept}). 
The value of $A_{\uparrow\downarrow}$ should depend on the absolute value of $B'$ as well as on the 
angle $\beta$ between the ordinary and mirror fields.

Following Ref.~\cite{Berezhiani:2008bc} we can then derive the oscillation time
\begin{equation}
    \tau_{nn'}(B, B') = \sqrt{\frac{t_\text{s}}{\langle t_\text{f} \rangle A_{\uparrow\downarrow}}}\sqrt{f_A(B,B')\text{cos}\beta},
\label{eq:taunnprime}
\end{equation}
where $\langle t_\text{f} \rangle$ is the average value of 
the mean free flight time of the UCN during storage,
$f_A(B,B')$ 
a simple function of $B$ and $B'$~\cite{Abel:2020kdg},
$\beta$ the angle between the vectors $B$ and $B'$.
Obviously, in the absence of an effect, the asymmetry would be zero and a limit on  
$\tau_{nn'}$ can be derived, 
albeit in combination with the dependence on the 
angle $\beta$ between $B$ and $B'$.

Also an 
asymmetry $E$ can be measured, by comparing  
the UCN counts averaged between $+B$ and $-B$ cycles 
with 
the counts $n_0$ measured at $B=0$, with
\begin{equation}
E \equiv \frac{2 \ n_0}{n_\uparrow - n_\downarrow} -1 = \frac{t_s}{\langle t_f \rangle} \frac{1}{\tau_{nn'}^2} f_E(B,B')
\label{eq:Easymmetry}
\end{equation}
with $f_E(B,B')$ is another simple function of $B$ and $B'$~\cite{Abel:2020kdg}, 
but different from $f_A$.

The latter asymmetry should not depend on the unknown 
direction of the mirror field 
but only on its absolute value.  
Namely, by measuring the values of $E$ and $A$ consistent with zero
at different 
applied fields $B$, one sets the limits on
respectively $\tau_{nn'}$ and $\tau_{nn'}/\sqrt{\cos\beta}$ 
for a given inferred value of $B'$. 
In a more general case when both $n-n'$ and $n-\bar{n}'$ oscillations are present,   
e.g. $E$-measurements would restrict an effective combination 
$\tau_{\rm eff} = (\epsilon_{nn'}^2 + \epsilon_{n\bar n'}^2)^{1/2}$~ \cite{Berezhiani:2020vbe},   
and thus restrict $\tau_{nn'}$ and $\tau_{n\bar n'}$.  

As the possibility of non-zero $B'$ came into focus, 
experiments quickly measured $E$ and $A$ asymmetries at different values of $B$
excluding also a sizeable fraction of the parameter space 
with non-vanishing $B'$ 
fields~\cite{Altarev:2009tg,Bodek:2009zz,Serebrov:2009zz,Berezhiani:2017azg,Abel:2020kdg}. 
However, some measurements have shown deviations from the null-hypothesis.   
In particular, reanalysis of the data of the experiment~\cite{Serebrov:2009zz} 
in Ref.~\cite{Berezhiani:2012rq} revealed signal-like effects at the $5\sigma$ level. 
Ref.~\cite{Serebrov:2009zz} reports a $3\sigma$ deviation from zero 
for the $A$-asymmetry after averaging between measurements 
at different values of $B$,
which is problematic once the $n-n'$ oscillation probability 
is supposed to have a resonant dependence on $B'$ \cite{Berezhiani:2008bc}.  
In addition, the data of experiment~\cite{Berezhiani:2017jkn} have shown a
$2.5\sigma$ deviation, again for the $A$-asymmetry.

\begin{figure}[htb]
\begin{center}
%
\resizebox{0.73\textwidth}{!}{\includegraphics{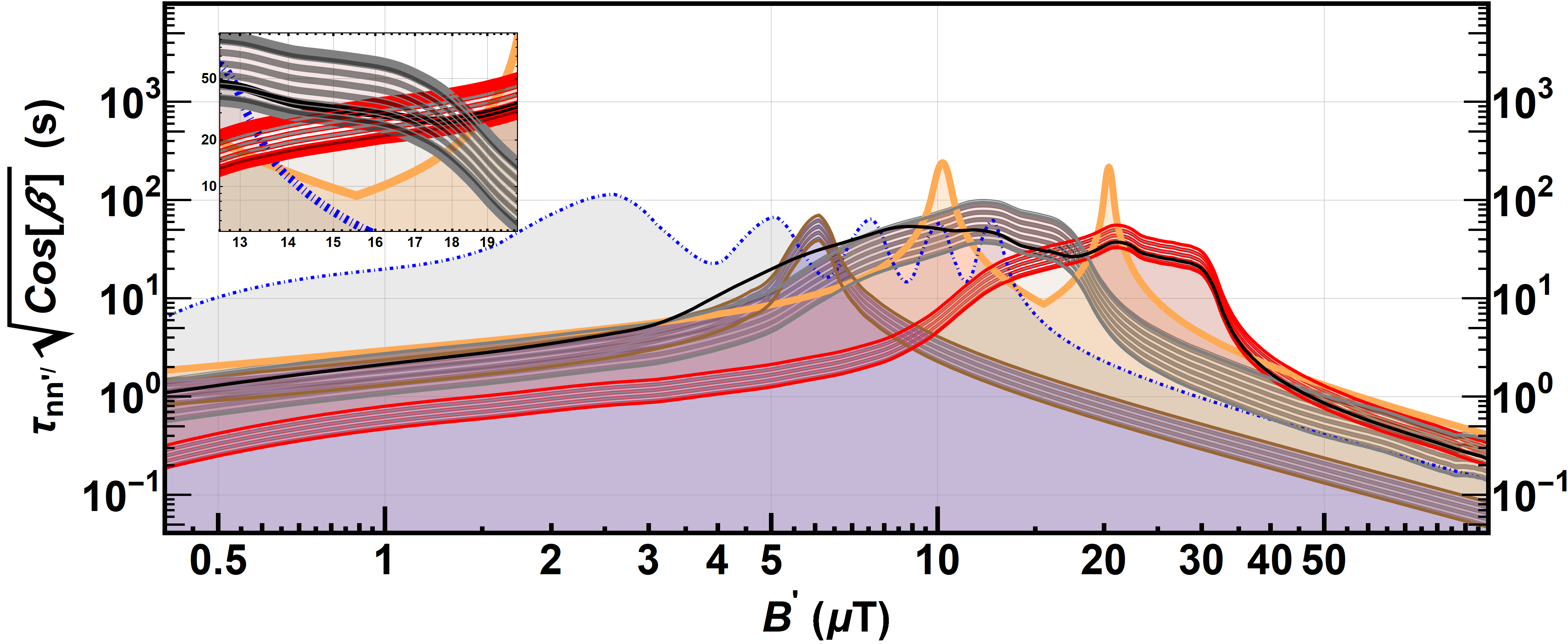}}
\caption[$nn'$ limits]{Limit on $n-n'$ oscillation times normalized by 
the square root of the cosine 
of the angle $\beta$ 
between the magnetic field $B$
in vertical direction and the unknown field $B'$.
The figure is
from the recent publication  
of the nEDM collaboration~\cite{Abel:2020kdg}.
White regions are still allowed.
The solid orange curve represents the lower limit from~\cite{Abel:2020kdg}.
The solid black curve is the global constraint 
calculated in~\cite{Berezhiani:2017jkn} 
considering data
from~\cite{Ban:2007tp,Serebrov:2007gw,Altarev:2009tg,Serebrov:2009zz,Berezhiani:2017jkn}. 
The dashed blue curve is from~\cite{Altarev:2009tg}. 
The red region 
bounded by 
red lines 
peaking above 20\,$\micro$T 
is the 95\% C.L. signal
region recalculated in Ref.~\cite{Berezhiani:2017jkn} 
from the 5.2$\sigma$ deviation observed in Ref.~\cite{Berezhiani:2012rq} 
in the experimental data of Ref.~\cite{Serebrov2008mirror}.
The brown region bounded by 
brown lines peaking at around 
6\,$\micro$T is the 95\% C.L. signal region calculated
by the same Ref.~\cite{Berezhiani:2017jkn} 
from a 3$\sigma$ deviation in~\cite{Ban:2007tp}. 
The gray region bounded by dashed gray lines indicates 
the 95\% C.L. region for the 2.5$\sigma$ deviation observed 
in the B2 series of experiment~\cite{Berezhiani:2017jkn}. 
%
The angle  of 2.9$^\circ$ between the vertical axes 
at the geographic locations of PSI and ILL 
introduces a small additional source of uncertainty when comparing exclusion plots
from measurements at PSI and ILL, respectively. 
Reproduced with permission under the Creative Commons CC-BY license. 
Indicated changes are: The
upper part of the figure was omitted in this reproduction and the citations
given in the figure removed. 
}
\label{fig:nnp_asym_lim}
\end{center}
\end{figure}

Using the nEDM spectrometer the nEDM collaboration at PSI performed 
a $n-n'$ search in 2017 which was designed to test the deviations and  
allowed signal regions 
claimed in Ref.~\cite{Berezhiani:2012rq}. 
%
%
%
%
The exclusion plot shown in Fig.~\ref{fig:nnp_asym_lim}
summarizes the status of $n-n'$ experimental results from the 
asymmetry channel.
%
%
The claimed potential signals of~\cite{Berezhiani:2012rq}
are excluded by the 2017 data of the nEDM 
collaboration
analyzed 
in the context of a PhD thesis~\cite{Prajwal2019}
and recently published~\cite{Abel:2020kdg}. 
However, further reanalysis~\cite{Berezhiani:2017jkn} of the 
magnetic field conditions of 
the experiment~\cite{Serebrov:2009zz}
resulted in a wider range of possible signal regions indicated by 
the signal bands in the figure. 
These new potential signal regions were only partially excluded 
as this information was not yet available at 
the time of planning and data taking 
in 2017.

\section{Experiment}
\label{Sec:Experiment}
%

%

We present in this work a novel experiment designed to probe the entire relevant parameter space, and to 
confirm or refute the claimed mirror neutron signals compatible with previous 
results. 
In 2020 the first version of the apparatus was used in a short beam time to demonstrate 
its performance at the West-1 beamport
of the ultracold neutron source at PSI~\cite{Lauss2012,Lauss2012b,Lauss2014,Bison2017,Bison2020,Lauss2021}.
In spring 2021
several aspects of the apparatus were improved in preparation 
for the main data-taking run.

\subsection{Concept}
\label{subsec:Concept}

The concept of this experiment is based on
the search for increased losses of ultracold neutrons 
during storage,
in resonance with a vertical magnetic-field at specific field values.
%
%
In our measurements the number of UCNs after storage 
are normalized 
to the initial UCN density. 
If there would be neutron to mirror-neutron oscillations 
during storage enhanced in resonance either at $+B$ or $-B$, 
the asymmetry $A_{\uparrow\downarrow}$ 
(see Eq.~\ref{eq:asymmetry})
would be 
different from zero.

Our experiment allows two modes of operation, storage and leakage mode.
\begin{enumerate}

\item Classical storage measurement (sketched in Fig.~\ref{fig:cycle}): 
UCNs are filled into a storage volume. To measure the initial UCN density, 
the shutter to the detector SH2 is opened for the 
last ten seconds of the filling (``monitoring'') then closed.
The filling shutter SH1 is closed, and UCNs are stored for a given storage time. 
Then, the second shutter SH2 opens
and UCNs are emptied and counted in a detector (``counting''). 
The time spectrum of UCNs arriving in the detector during such a typical cycle is shown in Fig.\ref{fig:UCNTimeSpectrumSingle}.

\item Leakage measurement (sketched in Fig.~\ref{fig:cycle-leak}): 
UCNs are filled into the storage volume
and the filling shutter SH1 is closed,
but SH2 and hence the storage volume towards the detector
are permanently open. 
%
An additional shutter suspended from the top (a modification of SH3 which reaches to the bottom of the storage vessel) allows to have a small, adjustable opening to the detector.
Hence UCNs are counted continuously. 
%
The measured UCN rate at the detector during such a cycle 
is illustrated in Fig.~\ref{fig:UCNTimeSpectrumLeakage}.
In this scheme the experiment is not only sensitive to 
variations in total counts but also to the change of the leakage time spectrum.

\end{enumerate}

\begin{figure}[htb]
	\centering
	 \includegraphics[width=0.70\textwidth]{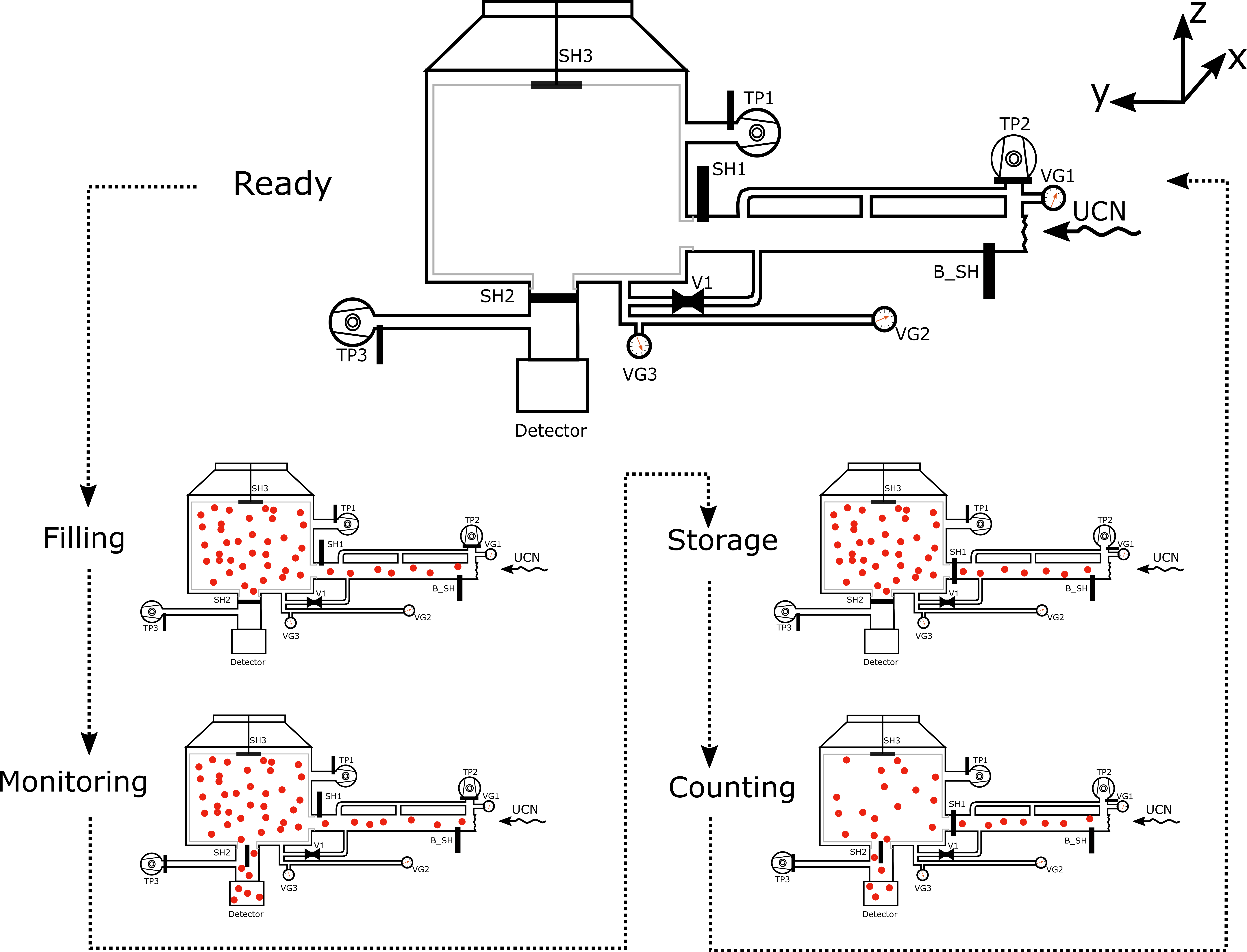}
	\caption[Illustration of steps of a ``storage'' type measurement]{
		Illustration of steps of a ``storage'' type measurement.
		Indicated in the figure are turbo-molecular pumps (TP1, TP2, TP3),
		shutters (SH1, SH2, SH3, B\_SH),
    vacuum gauges (VG1, VG2, VG3),
		and valve (V1). 
		(Not to scale).		  
		The operation sequence is the following:
		- Ready: Setup is waiting for the beam pulse and pumping.
		- Filling: UCNs enter the storage vessel with SH2 and SH3 closed.
		- Monitoring: After filling a short monitoring period, with SH2 open, 
		is used to determine the UCN intensity.
		- Storage: SH1 and SH2 is closed and UCNs are stored for a set time if 120\,s in the storage vessel.
		- Counting: SH2 is opened and UCNs are counted in the detector.	
	}
	\label{fig:cycle}
\end{figure}

\begin{figure}[htb]
	\begin{center}
		\resizebox{0.6\textwidth}{!}{\includegraphics{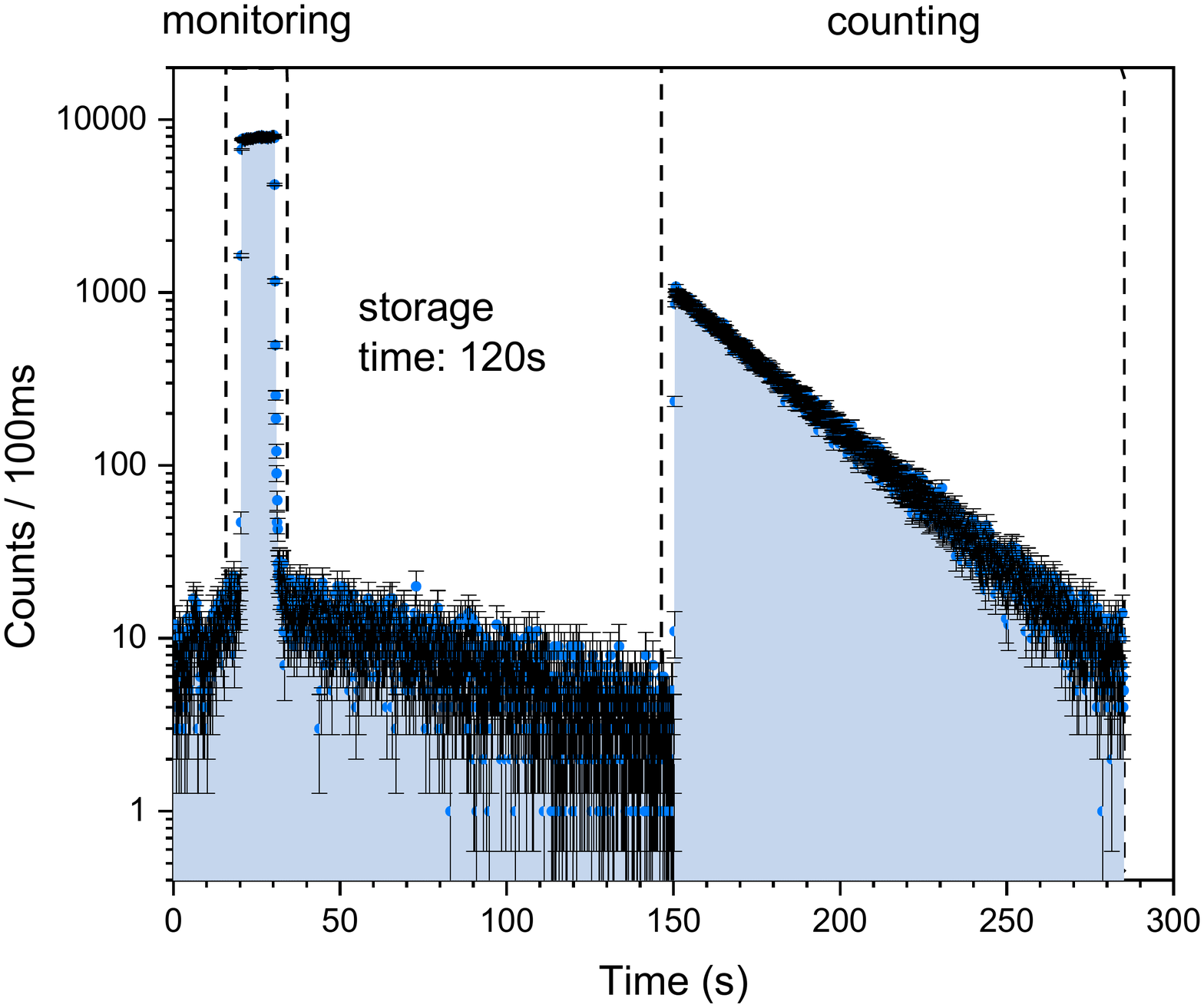}}
		\caption[``Storage'' type measurement cycle time spectrum]{
				Time spectrum of UCN counts in the UCN detector for a single ``storage'' type measurement.
		}
		\label{fig:UCNTimeSpectrumSingle}
	\end{center}
\end{figure}

\begin{figure}[htb]
	\centering
	\includegraphics[width=0.70\textwidth]{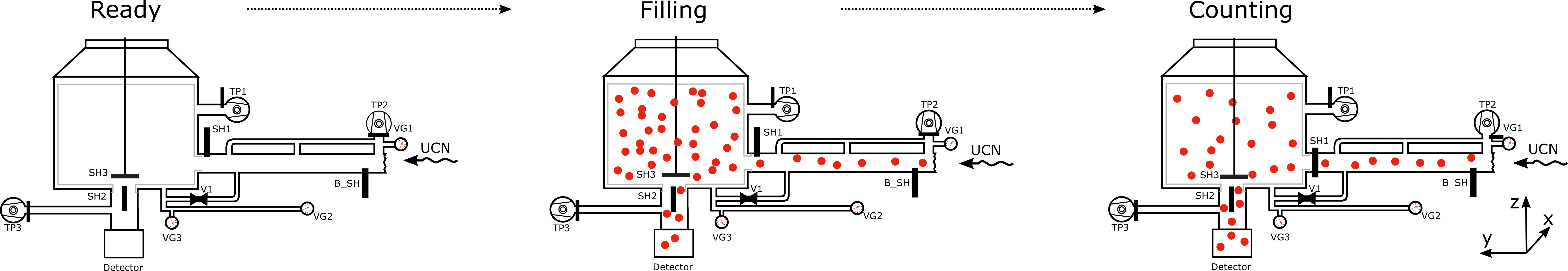}
	\caption[Illustration of steps of a ``leakage'' type measurement]{
		Illustration of steps of a ``leakage'' type measurement. 
		Shutter SH3 was lowered to the bottom of the storage vessel allowing thus to define
		a specific leakage rate, and the top 
		opening (for pumping) was closed with an additional plate.
		The operating procedure is:
		- Ready: The setup is waiting for the beam pulse and pumping.
		- Filling: UCNs are filled into the storage vessel and the emptying guide
		up to the detector (SH2 open).
		- Counting: Shutter SH1 is closed and all UCN are stored in the storage vessel and
		counted simultaneously in the detector (SH2 open).
	}
	\label{fig:cycle-leak}
\end{figure}

\begin{figure}[htb]
	\begin{center}
		\resizebox{0.5\textwidth}{!}{\includegraphics{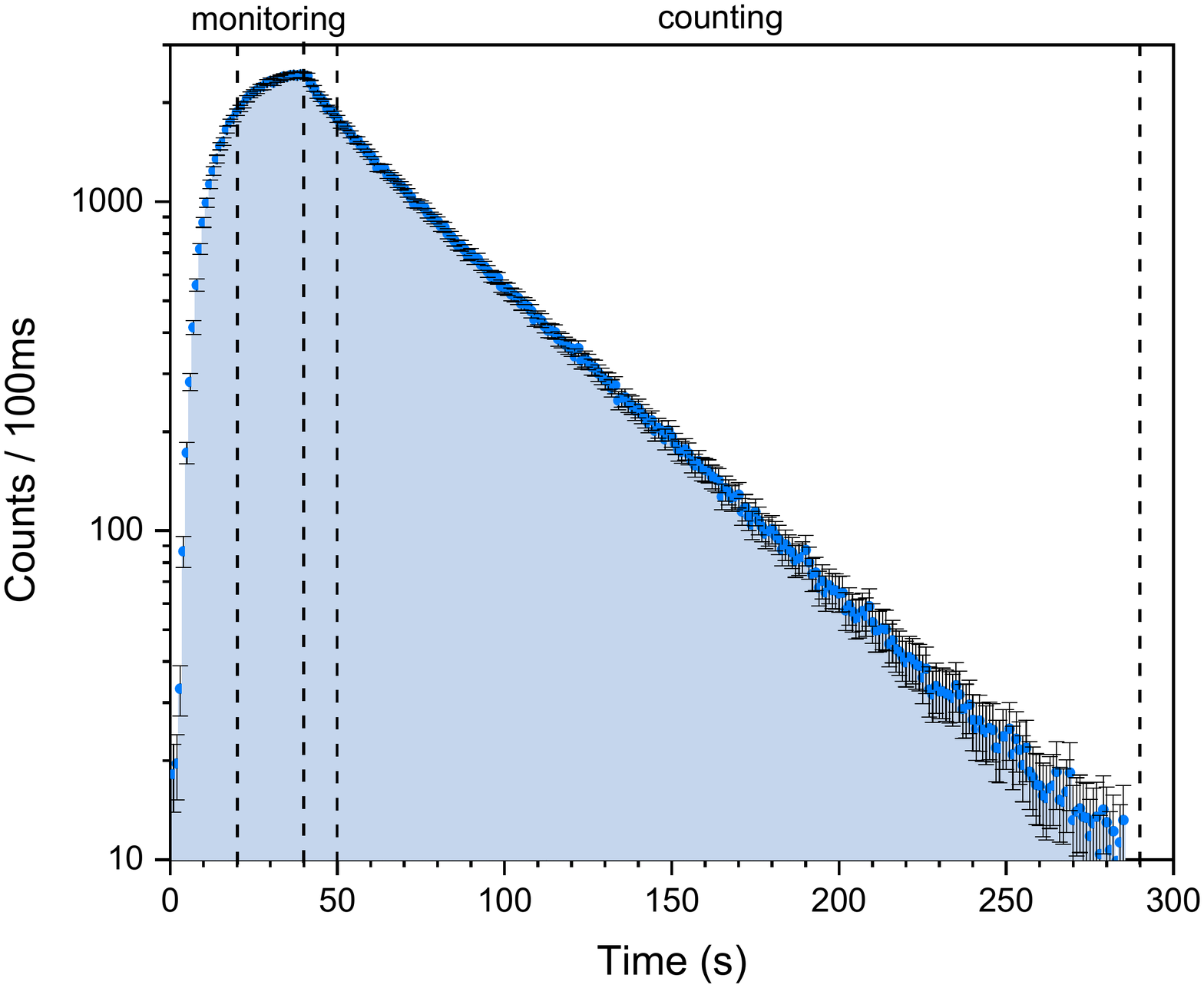}}
		\caption[``Leakage'' type measurement cycle time spectrum]{
			Time spectrum of UCN counts in the UCN detector for a single ``leakage'' type measurement.
		}
		\label{fig:UCNTimeSpectrumLeakage}
	\end{center}
\end{figure}

\subsection{Ultracold neutron system}

The setup of the experiment can be explained 
following
the cut-view CAD drawing shown in Fig.~\ref{fig:Exp-CAD}
as used for the fall 2020 measurements.
UCNs are guided over 2750\,mm from the West-1 beamport shutter to the 
entrance shutter of the storage vessel using
nickel-molybdenum 
(NiMo) coated glass guides \cite{Daum2014,Blau2016,Bison2020}
with an inside diameter of 130\,mm.
The 1.47\,m$^3$ large UCN storage volume placed inside a large vacuum 
vessel is made of non-magnetic stainless steel 
and
extends via 200\,mm diameter stainless steel tubes
to two shutters.
The filling shutter SH1 is a VAT product as described in Ref.~\cite{Blau2016}.
It is coated with diamond-like carbon and vacuum tight.
The emptying shutter is from the standard bottle setup~\cite{Bison2016}
and made from stainless steel.
It has a small ($\sim$1\%) UCN leakage, but opens and closes fast 
and reproducibly. 
The top of the volume has a 200\,mm diameter hole
which can be opened during pumping periods to
increase pumping efficiency.
It is closed during UCN storage, otherwise a large 
fraction of UCNs is lost.
The vacuum vessel is made from stainless steel and 
all parts of the vessel were
checked for possible magnetic contamination 
and replaced where possible.

The detector, a 
Cascade UCN detector\footnote{CDT Cascade Detector Technologies, 69123 Heidelberg, 
Germany, cd-t.com} 
with a sensitive area of
20\,cm$\times$20\,cm 
is located below the emptying shutter and connected via
a short stainless-steel guide section 
which can be evacuated via a 5\,mm opening.

The magnetic field is generated via a 3-dimensional coil system
as described and first used in~\cite{Brys2005}.
The photo in Fig.~\ref{fig:photo-setup}
shows the setup as it was installed and
used for the test measurements in fall 2020
in UCN area West-1.

\begin{figure}[htb]
\centering
\includegraphics[scale=0.45]{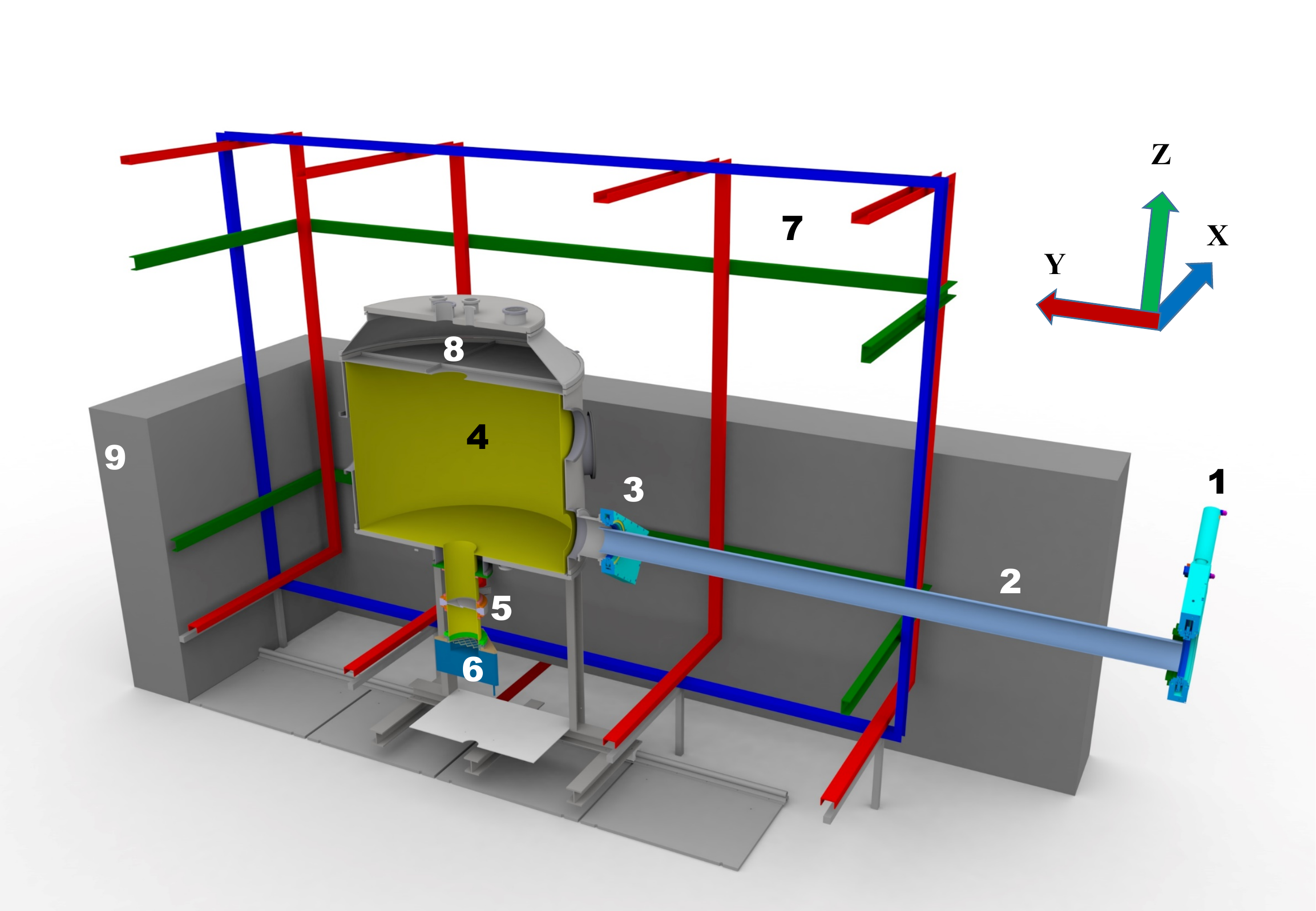}
\caption[Cut-view of experiment]{
Cut view of the experiment as installed in area West-1 
in fall 2020:
1) beamport shutter,
2) UCN guides (275\,cm),
3) storage vessel (filling) shutter SH1,
4) vacuum vessel containing the storage chamber,
5) storage vessel (emptying) shutter SH2,
6) UCN detector,
7) coil system for B-field generation in direction
x: blue, y: red, z: green,
8) top hole opened during pumping (can be closed by shutter SH3 in storage measurements),
9) concrete shielding.
}
\label{fig:Exp-CAD}
\end{figure}

\begin{figure}[htb]
\centering
\includegraphics[scale=0.15]{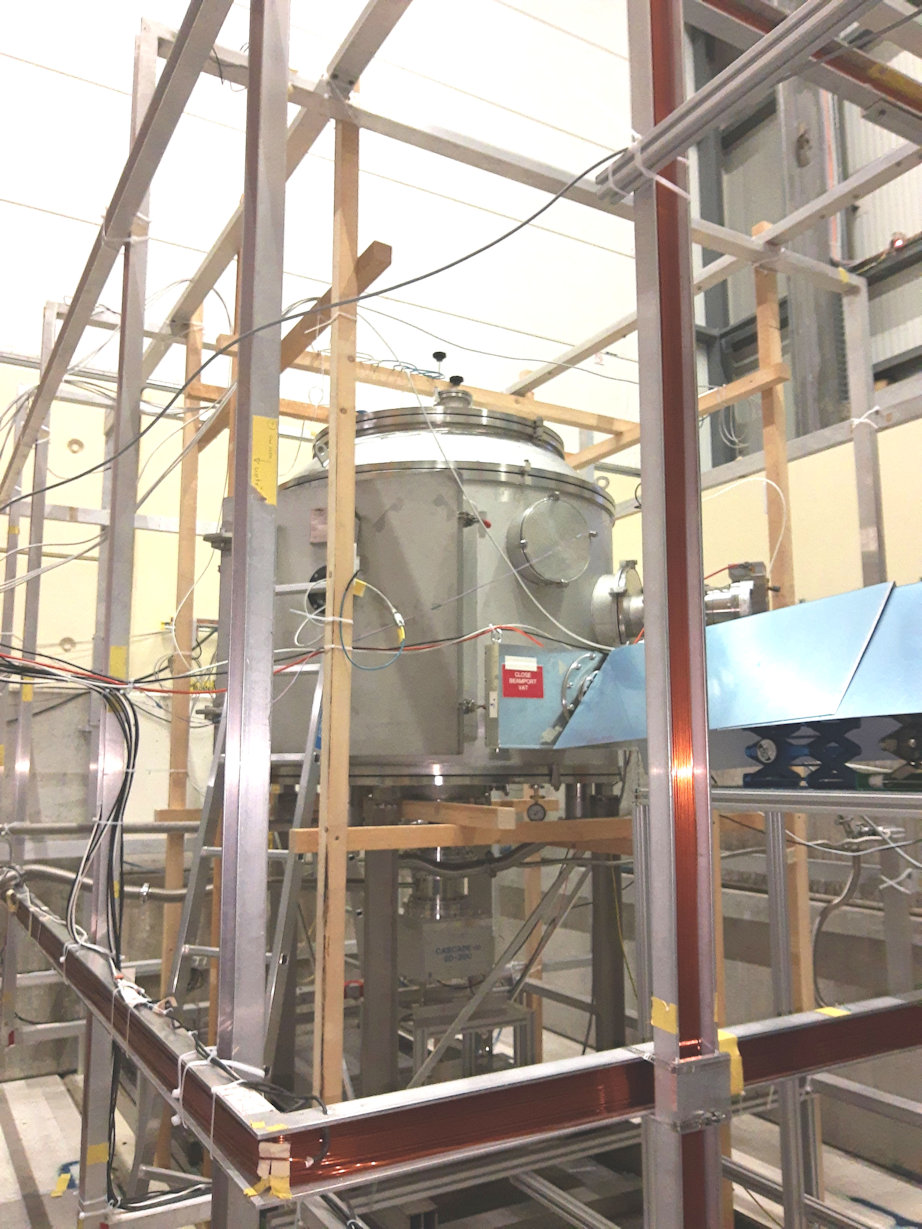}
\caption[Setup in fall 2020]{
Photo of the experiment setup in the UCN area West
during the test measurements in fall 2020.
The visible parts can be compared with the description
given in Fig.~\ref{fig:Exp-CAD}.
The UCN glass guides (on the right hand side)
leading from beamport West-1
to the apparatus are covered by 
folded sheet metal pieces for protection.
The visible metal frame supports the coil-windings, 
the wooden frame supports the fluxgates.
}
\label{fig:photo-setup}
\end{figure}

\subsection{Improvements of the ultracold neutron system in 2021}
\label{subsec:UCNupgrades}

During the measurements in fall 2020 we have identified various
parts of the UCN system which could be slightly upgraded to improve the performance.
These modifications will increase the
UCN statistics and the UCN storage time constant. 
Each of these changes has been found to substantially improve 
the achievable statistical sensitivity in simulations (see Sec.~\ref{sec:upgradesim}).
Some modifications were already implemented and tested:
\begin{itemize}
	
	\item {In order to increase the UCN storage properties
		of the storage volume, 
		the entire stainless-steel
		body of the vessel was electro-polished.
		This increased the storage time constant
		as the entire surface and especially the welding seams 
		became cleaner.
	}

	\item {The UCN guide on the bottom
		of the storage vessel towards the detector
		is made from stainless steel with a Fermi potential
		$V_F$=185\,neV.
		Coating of the surface with a material of higher $V_F$,
		namely NiMo with a
		$V_F$=220\,neV \cite{Blau2016} was applied
 	 in order to further reduce wall losses in this region. 
	}
	
	\item {New UCN guides from the beamport to the storage vessel,
		with a ID=180\,mm NiMo-coated glass guides
		and a ID=200\,mm polished stainless-steel guide, 
		effectively doubled
		the cross-section of the guides, however, with the drawback of the
		metal guide part having a larger surface roughness than glass.
		Still, this resulted in an increase in the UCN filling rate.
	}

	\item {A new plate shutter 
	(a dynamic version of SH3, shown in Fig.~\ref{fig:cycle-leak}, 
	that would open and close the storage volume to the detector)
	was designed and tested. At the same time the shutter SH2 of the standard bottle
	used in 2020 was coated with NiMo and found to be superior.}
	
	\item {With our UCN simulation tuned to the further detailed measurements,
		we re-investigate changing the height of the
		storage vessel with respect to the beamport.
	}
	
\end{itemize}

\clearpage

\subsection{Magnetic field system}
\label{subsec:magn}
                                                
To compensate ambient magnetic fields and to produce the desired field for the 
experiment, a set of 
rectangular coil pairs
were erected around the 
experiment (see Figs.~\ref{fig:Exp-CAD} and~\ref{fig:photo-setup}), 
with the neutron storage chamber approximately at the center. 
The coil system described in~\cite{Brys2005} has been recommissioned. 
It has nominal dimensions of 3\,m $\times$ 3\,m $\times$ 4.6\,m. 
In the $x$ and $z$ directions, the field is produced by pairs 
of 3\,m $\times$ 4.6~m rectangular coils, separated by around 1.8\,m. 
In the $y$ direction, a set of four square 3~m $\times$ 3\,m coils are located 
at $y=\pm0.9$\,m and $y=\pm2.2$\,m, with the central coils having 
four times as many turns as the outer coils. 
Additionally, the current of each coil can be individually controlled, 
allowing 
an adjustment of field gradients.

The homogeneity H(B) of the magnetic field produced by the $B_z$ coil was 
computed over the 
neutron storage volume based on the Biot-Savart law,
using 
\begin{equation} 
H(B)= ( B_z^i - B_z^{mean} ) / B_z^{mean}  
\end{equation}
with $B_z^i$ the magnetic field computed at several positions 
covering the UCN storage vessel, 
and their mean value $B_z^{mean}$.
The standard deviation $B_z^i - B_z^{mean}$
is illustrated in Fig.~\ref{fig:NickKickOffSlide10}.
The resulting H(B)<0.5\% 
is considered to be good enough
that it does not limit the quality of the field within the chamber. 
This is ultimately limited by local field inhomogeneities 
and environmental fields, such as those produced by vacuum pumps and 
other components critical to the experiment, 
and those produced by the 
large steel-armed concrete blocks forming the floor and the walls 
of the experimental area.
To minimize local inhomogeneities, magnetic components such as certain 
vacuum gauges and pumps
were all moved with extension tubes
outside the coil cage for the main run in 2021.

\begin{figure}[htb]
	
	\centering
	
	\includegraphics[scale=0.7]{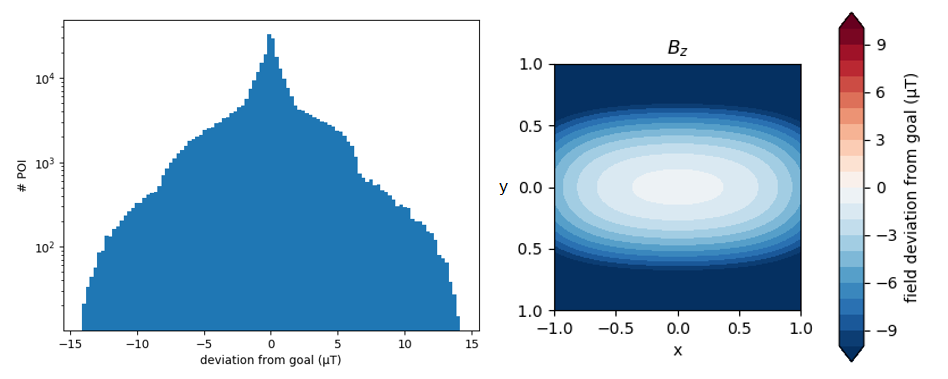}
	
	\caption[Simulated magnetic field map at based on a coil 
	configuration before full optimization]{
		Simulated magnetic field map based on a coil configuration before 
		full optimization. The simulation is based on an analytical calculation 
		using the Biot-Savart law, with coil currents set to produce a 
		nominal field of \SI{100}{\micro T}. 
		The left plot shows a histogram of $B_z - B_{z}^{mean}$ at a 
		large number of points distributed randomly throughout 
		the cylindrical storage volume (note the logarithmic y-axis). 
		The right plot shows a slice of the computed field deviation in the $xy$ plane 
		cut at $z=0$ i.e. the center of the storage volume, with $x$ and $y$ in meters. 
		This calculation assumes linearity, so the deviations in microtesla from a 
		uniform \SI{100}{\micro T} field are equivalent to a percentage deviation from an 
		arbitrary field. The real apparatus used an optimized coil geometry, 
		however here we present calculations using a ``realistically bad'' 
		configuration to estimate the impact of uncompensated inhomogeneities.}
	
	\label{fig:NickKickOffSlide10}
	
\end{figure}

Many sources of magnetic field disturbances from other installations 
surrounding UCN Area West are known from the nEDM measurements
in UCN area South~\cite{Afach2014JAP}. 
In particular a large, nearby superconducting magnet is known to produce 
changes in the magnetic field of order \SI{100}{\micro T} when it is ramped.

During the measurement an active compensation of the ambient magnetic 
field was applied. 
In 2020 
a set of 13 Stefan Mayer FLC3-70 three-axis 
fluxgates was located in the vicinity of the vacuum tank. 
The locations were chosen 
to be able to optimally calculate the magnetic field in the storage tank 
and at the same time
to minimize the impact of possible local magnetic-field sources 
such as vacuum pumps. 
The range of $\pm$\SI{200}{\micro T} of the sensors 
set the maximum mirror field that 
could be probed in 2020. 
The improved setup in 2021 
includes ten 3-axis Sensys FGM3D/125 fluxgates to cover the field range up 
to 125\,$\micro$T with a higher accuracy, five 3-axis Stefan Mayer FLC3-70 
fluxgates for a consistency check with the previous year, 
and one higher range Sensys FGM3D/1000 sensor to allow us to probe 
even higher fields,
limited only to about $\pm$\SI{380}{\micro T} 
by the maximum current of the power supplies of the coils.
The fluxgates are read out at 50\,Hz by a 
computer-based data-acquisition (DAQ) system 
fully integrated into the main DAQ system (discussed in full in Sec.~\ref{sec:daq}).

The coils are powered by six computer-controlled DC power supplies each capable of 
supplying 10/20\,A at 35\,V, with the possibility to reverse the polarity provided by a system of relays. 
Two algorithms have been implemented for the control of the coil currents:
\begin{itemize}
	\item Dynamic: the currents in each coil are set according to a 
	modified version of the ``dynamic'' algorithm described in~\cite{Afach2014JAP}, 
	aiming to match the readings of each of the first 10 fluxgates to the goal field using a PI feedback algorithm.
	\item Static: the currents in each coil are set to values 
	to achieve the desired fields based on offline measurements, 
	assuming the external field remains constant.
\end{itemize}
When it is known that the external field may change, it is mandatory to use the dynamic mode.
The coil current is updated at 10\,Hz. This is more than enough to 
allow dynamic compensation of external field changes on the relevant timescale 
(namely, the average field change over the entire storage time), 
and even to allow the goal field to be changed during a live measurement. 
This allows simplification of the analysis by permitting 
the detuning of the magnetic field during the filling of 
the neutron storage vessel.

Based on the measured fields in each of the fluxgates, 
a first estimate was that the actual field homogeneity achieved is 
better than 3~$\micro$T at a field setting of 10 to 20~$\micro$T.
This is likely an overestimate, as the fluxgates are located 
outside the chamber and therefore 
influenced by ambient magnetic field sources. 

The effect of field inhomogeneity is 
a slight broadening 
of the range 
of mirror fields probed by each run, as effectively multiple field 
values are probed simultaneously. 
Our experiment is not targeting the $B'=0$ region and 
therefore does not require zero field capability on a level below a few $\micro$T. 
Better knowledge of the magnetic field was acquired during 
a dedicated mapping campaign before the 2021 measurement 
period (see Sec.~\ref{subsec:mapping})
with the full analysis not yet finished.



\subsection{Mapping of the magnetic field}
\label{subsec:mapping}

For the interpretation of the neutron storage measurements 
in terms of a mirror neutron 
signal, it is important to quantify the field distribution 
within the neutron storage 
chamber during measurement. 
During neutron data-taking, a set of 13 three-axis fluxgate 
magnetometers monitored in 2020 the field in the vicinity of the storage
chamber (see Sec.~\ref{subsec:magn}). 
However, after all magnetic apparatus components were moved
outside the coil cage, the remaining ambient field
in the proximity of the storage chamber due to non-removable parts
of the apparatus,
i.e. magnetized welding seams on flanges or on the floor of the vacuum tank, 
and the nearby armed concrete shielding,
makes a simple interpolation or fitting difficult.

A mapper device was designed and prototyped, inspired by the mappers 
used by the nEDM and n2EDM experiments \cite{Ayres2021Mapper,n2edm2021}, 
with substantial simplifications. 
A vertical axis passes through bearings mounted on 
flanges on the top and bottom of the vacuum tank. 
On this axis, a perpendicular arm is mounted within the 
storage chamber. By raising, lowering, or rotating the axle, 
the arm can be moved to sweep within the closed chamber 
from outside. On the arm, several fluxgate magnetometers 
are mounted at different radii. This allows the magnetic field 
to be probed in almost the entire volume directly. 

\begin{figure}[htb]
	
	\centering
	
	\includegraphics[width=0.35\textwidth]{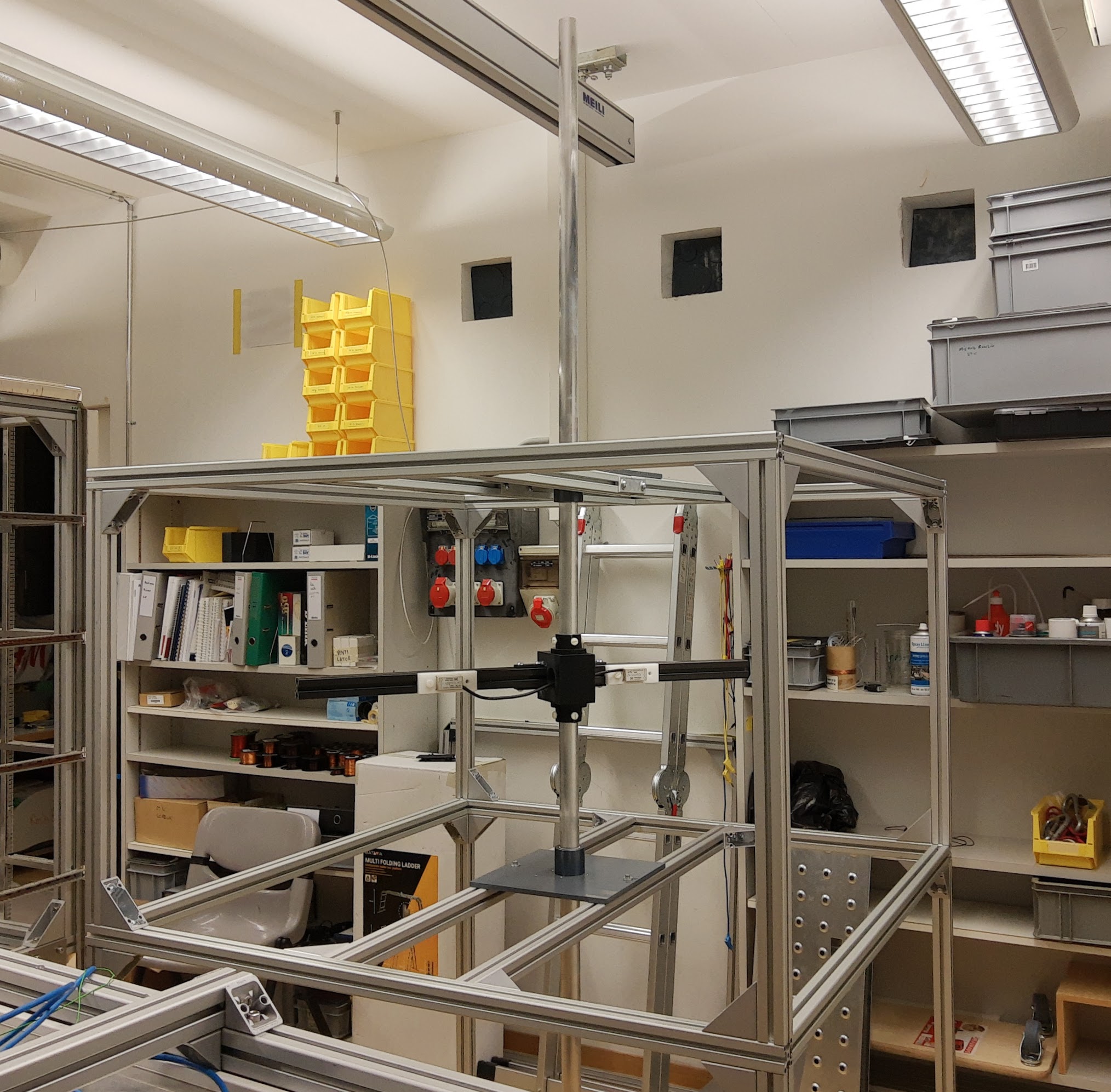}
	
	\caption[Mapper mounted in test stand]{
		Photo of the prototype magnetic-field mapper mounted in a test stand in the lab at ETH Zurich. The central axle can slide freely through flanges to be mounted on the top and bottom of the vacuum tank, allowing the height and rotation of the arm to be manipulated from outside with the tank closed. Several fluxgates can be mounted at different radii on the arm, allowing almost the full 3D volume to be mapped. All parts are non-magnetic.}
	
	\label{fig:mapper}
	
\end{figure}

The mapper was constructed within a test frame in 2020 
(photographed in Fig.~\ref{fig:mapper}), 
and finally commissioned and installed in the 
apparatus in spring 2021. 
The measurements were completed within several weeks and are currently being
analyzed.

\subsection{Data acquisition system}
\label{sec:daq}

In order to reliably control the individual parts of the experiment and
to precisely repeat identical timing sequences, we developed an automated computer-controlled data acquisition and control system (DAQ). 

As the DAQ system for the upcoming n2EDM experiment~\cite{n2edm2021} 
is already well 
advanced in its implementation, and there is significant overlap
in the required functionality, it was decided to build the system 
up on the already existing n2EDM DAQ and infrastructure.

Notable simplifications from the full n2EDM system include:

\begin{itemize}
 \item relaxed timing: synchronization via NTP (millisecond precision) sufficient
 \item data generation limited: no fast storage backend necessary
 \item standalone UCN detector with separate DAQ
 \item simplified on-line analysis: no cycle-to-cycle information exchange
\end{itemize}

All systems are time-synchronized to a GPS-disciplined 
high-precision time server.
The central services of the DAQ system are running on a dedicated server
in the n2EDM counting room, with four separate modules connected through 
TCP/IP: 
\begin{itemize}
 \item Trigger: connected to the trigger signals sent by the proton accelerator HIPA to the UCN source in preparation of beam pulses.
 \item Shutters: control of the upstream and downstream UCN shutters (SH1, SH2),
both pneumatically actuated, via a Beckhoff EtherCAT control and 
read-back module\footnote{Ethercat is an Ethernet fieldbus standard 
communication system by 
Beckhoff Automation GmbH \& Co. KG, 33415 Verl, 
Germany, www.beckhoff.com}.
\item Coils: control and feedback of the magnetic-field generation system
(see Sec.\ref{subsec:magn}), also via Beckhoff EtherCAT modules.
\item UCN Detector: a standalone commercial CASCADE detector, triggered synchronously from the HIPA UCN trigger signals. Communication with the DAQ system via HTTP requests.
\end{itemize}

Each module was connected through a \emph{communication handler} program
to the central message distributor on the main server, where the system's
actions and schedules were defined in the \emph{sequencer}.
Pre-written sequences were loaded into the \emph{sequencer} as text files.
Logging and state information output as well as remote control 
was handled via a text terminal.
All data and log-files were written to
a common location on the main server.

Over the course of the test beamtime, approximately 6000 experiment
cycles were controlled and recorded by the DAQ system.
All central services of the DAQ system worked almost flawlessly throughout
the beamtime and minor bugs could be corrected early on.

\subsection{Supplementary measurements with neutrons}
\label{subsec:systematics}

One of the systematic effects influencing the result is related to 
the simulation of the average value of the 
free-flight time between wall collisions, 
$\left< t_f \right>$, which is a function of 
the UCN density and velocity distributions in 
and the exact geometry of 
the UCN storage volume. 
Up to now, the velocity information is extracted from simulations and 
fits to the measured UCN storage curves. 
We therefore plan two independent supplemental measurements:
\begin{enumerate}
\item Measurement of the UCN velocity distribution using 
the oscillating detector ``OTUS'', 
as described in~\cite{Rozpedzik2019},
developed at Jagiellonian University, Cracow.
\item Studies of the evolution of the UCN density distribution 
in the storage vessel using an endoscopic UCN detector 
with $E_F \leq 0$,
which is currently
under development at the University of Mainz. 
\end{enumerate}

\subsection{Demonstrated performance}

In 2020, we completed a test beamtime at 
the PSI UCN West-1 beamport. 
The average performance for the most important parameters are
given in Tab.~\ref{tab:performance}.
During this measurement period we aimed to
\begin{itemize}
	\item test the UCN properties of the guides, storage chamber and shutters,
	\item evaluate the effectiveness of the proposed strategies 
	to normalize nonstatistical fluctuations in the UCN source output, and
	\item complete realistic physics data-taking at some of the 
	most well-motivated mirror-magnetic-field values.
\end{itemize}

	\begin{table}
		\centering
		\begin{tabular}{|l | c|} 
			\hline
			\textbf{Performance parameters} & Fall 2020\\
			\hline\hline
			Average UCN counts during monitoring $C_M$ & $0.6\cdot10^6$ \\
			Average UCN counts after storage  $C_S$ & $0.2\cdot10^6$\\ 
			\hline
			Storage time $t_s$& 120\,s\\
			Simulated UCN mean free flight time <$t_f$> & 0.16\,s\\
			\hline
			Storage curve parameters & $\tau\textsubscript{1} = 45$\,s   \ \ \    $A\textsubscript{1} = 0.90$\\
			$A\textsubscript{1} \ e^{-t/\tau\textsubscript{1}} + A\textsubscript{2} \ e^{-t/\tau\textsubscript{2}}$ & $\tau\textsubscript{2} = 85$\,s   \ \ \    $A\textsubscript{2} = 1.14$\\
			\hline
			UCN pulse duration & 8\,s \\
			UCN pulse period   & 300\,s \\
			\hline
			Average proton beam current & 2.0 mA\\
			\hline
		\end{tabular}
		\caption{Average performance parameters during the 2020 beam period of
		the apparatus, the UCN source, and simulated setup parameters.
		}
		\label{tab:performance}
	\end{table}

The first task was to validate that the storage vessel could 
store neutrons with low losses on the vessel 
walls and no large leaks, and to characterize its performance. 
After some minor interventions, the performance was 
sufficient in terms of UCN statistics (see~Tab.\ref{tab:performance})
to allow the next steps of the 2020 measurement campaign.
We measured the
UCN storage curve, i.e. the normalized number of integrated counts ($C_S/C_M$ 
as a function of storage time.
The measured storage curve is presented in Fig.~\ref{fig:StorageCurveShutter}. 
The storage time was scanned repeatedly between \SI{10}{s} and \SI{360}{s} 
over a continuous 48 hour period while the remaining parameters remained unchanged.

\begin{figure}[htb]
	\begin{center}
		\resizebox{0.75\textwidth}{!}{\includegraphics{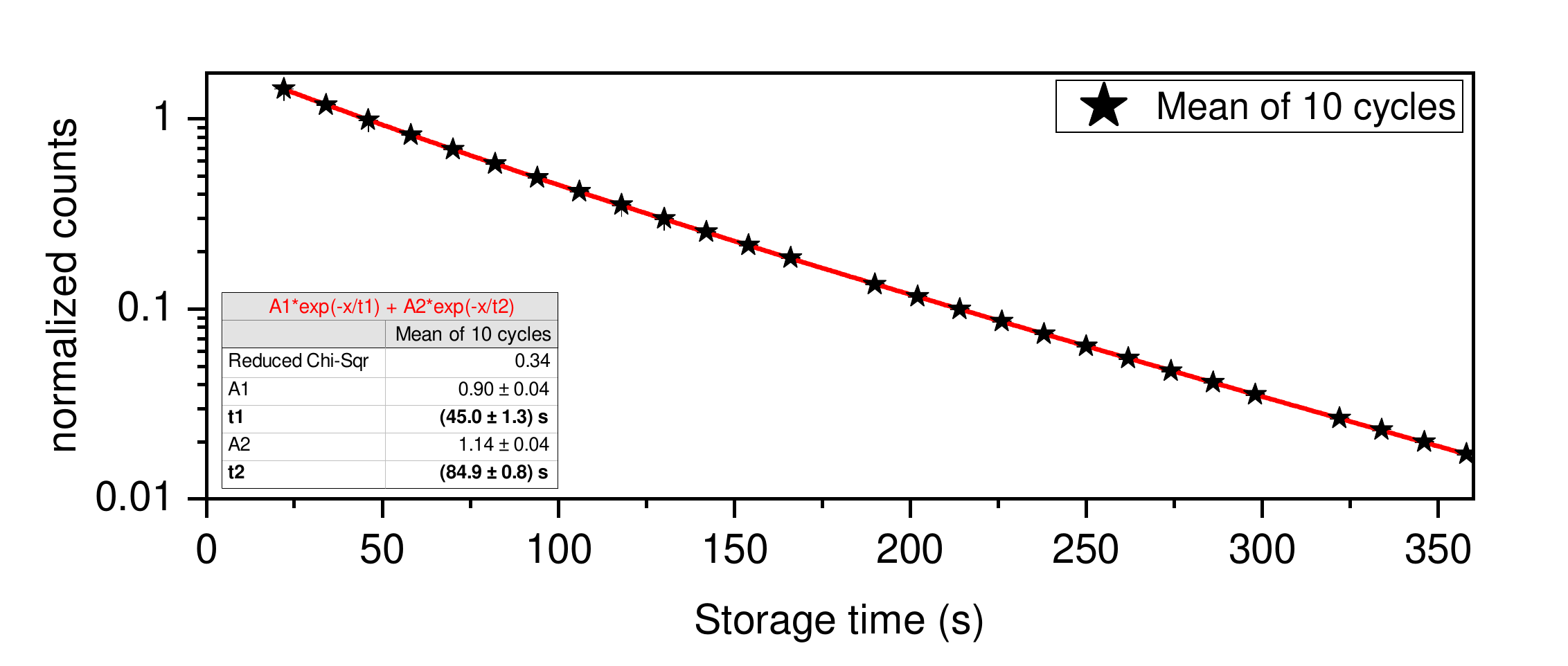}}
		\caption[Storage Curve]{
			Measured storage curve using the ``storage'' technique. 
			The data-points represent the mean of the normalized UCN 
			counts measured in ten measurement cycles 
			(i.e. as explained in Fig.\ref{fig:cycle}) per point. 
			Error bars are smaller than the symbols.
The found 'long' storage time constant of 85\,s was sufficient to 
propose the experiment at PSI, however, it was improved 
considerably due to the electropolishing 
for the 2021 data taking campaign
(see Sec.~\ref{subsec:UCNupgrades}).
Subsequent tests yielded a time 
constant with a preliminary value of 202\,s.}
\label{fig:StorageCurveShutter}
	\end{center}
\end{figure}

Following this, we aimed to demonstrate the monitoring 
methodology proposed to compensate for the
slow decrease in the UCN source output~\cite{Anghel2018}. 
We present one example measurement of the drift in 
monitoring counts $C_M$ 
and counts after storage $C_S$ over around a day 
and a half. 
The slow decrease in count rate occurs due to 
degradation of the surface of the solid deuterium in the 
source~\cite{Anghel2018} and other variations occur due 
to fluctuations in the accelerator and source operation. 
As part of the normal 
operation of the UCN source, a ``conditioning'' process 
where the surface is repaired, is performed several 
times per week, recovering the initial UCN output. 
The normalized count rate increases slightly over 
the measurement period after a conditioning. 
Typical behavior of the UCN counts 
following a conditioning is shown in Fig.~\ref{fig:RelativeCounts}.

\begin{figure}[htb]
	\begin{center}
		\resizebox{0.6\textwidth}{!}{\includegraphics{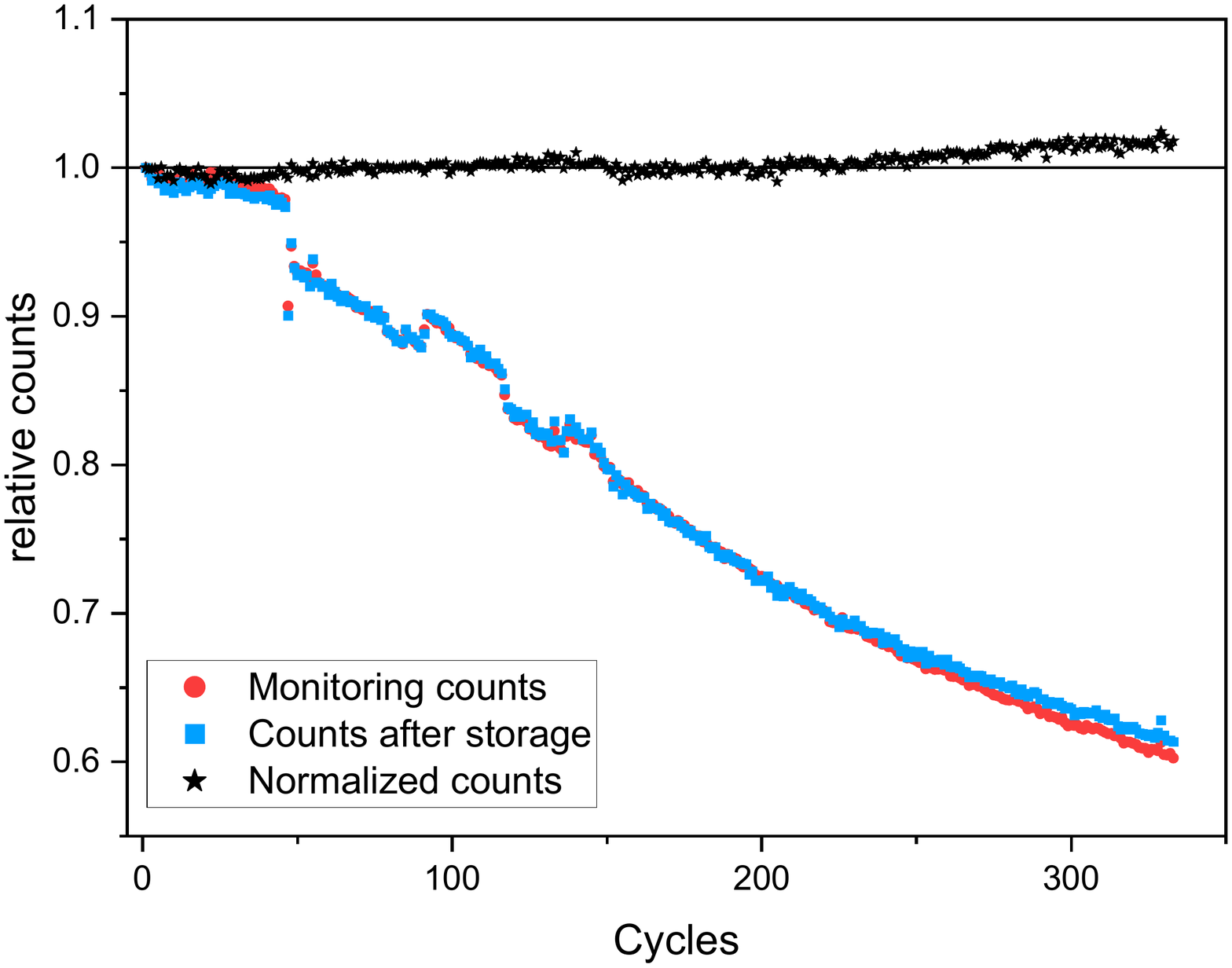}}
		\caption[UCN count variation during one measurement series]{
			Example measurement demonstrating the time variation 
			of the UCN counts over around one and a half days of normal running, 
			all normalized to 1 at the start.
			The length of one measurement cycle is given by the 
			5 minutes between two proton beam pulses onto the UCN source.
			Blue squares - integrated counts $C_S$ observed after storage;
			Red circles - integrated counts $C_M$ observed during the monitoring interval;
			Black stars - normalized counts $C_S$/$C_M$. 
		}
		\label{fig:RelativeCounts}
	\end{center}
\end{figure}

To compensate for remaining drifts, the polarity of the 
magnetic field is inverted in an eight-cycle 
sequence ``AB BA BA AB'', 
with, e.g., A denoting a cycle with the magnetic field pointing upwards 
and B a cycle with the field downwards.
Within each block, the asymmetry of the counts in the 
two antiparallel field configurations $\frac{A - B}{A + B}$ 
is computed. As illustrated 
in Fig.~\ref{fig:asymmetry0}, this ratio is found to be stable 
over two days, and compatible with zero, demonstrating the 
effectiveness of this scheme in compensating drifts. 
The scatter is found to be comparable to what 
would be expected for Poisson statistics, as shown in the 
right Fig.~\ref{fig:asymmetry0}.

\begin{figure}[htb]
	\begin{center}
		\resizebox{0.6\textwidth}{!}{\includegraphics{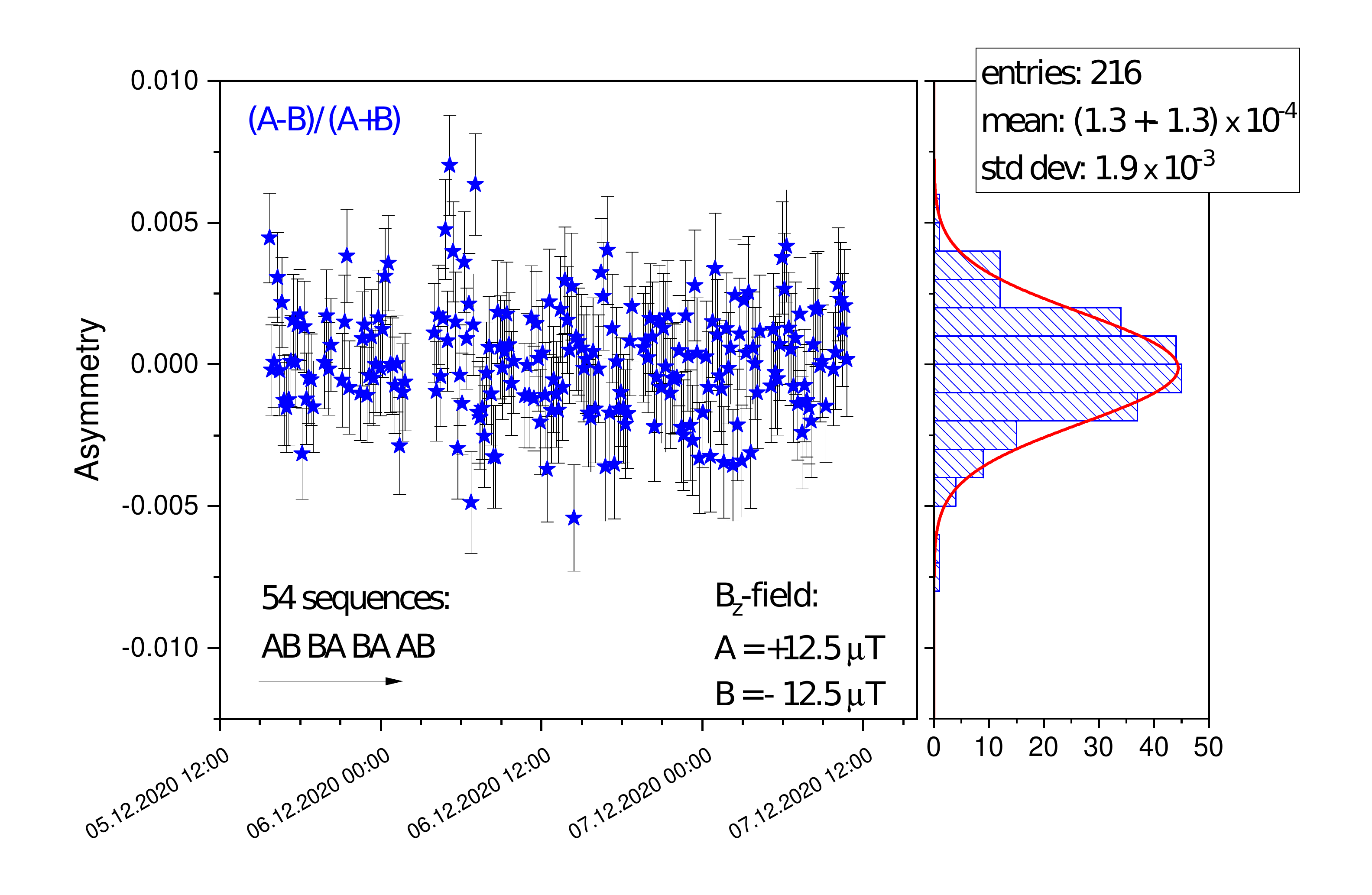}}
		\caption[Example asymmetry time series]{
			Representative time series of the count asymmetry computed for 
			54 sequences with opposite magnetic fields A and B, 
			as defined in the text (12.5$\micro$T for this example). 
			As shown on the right, the mean of the asymmetry values is compatible with zero 
			and the scatter is compatible to what would be expected for 
			pure Poisson statistics. 
			This demonstrates the suitability of our method to compensate 
			for small variations in the UCN source output.
		}
		\label{fig:asymmetry0}
	\end{center}
\end{figure}

During the 2020 beamtime, the leakage method was also investigated, 
where the lower shutter SH2 was permanently open, 
but with the lower hole blocked by a disk SH3 suspended from above
(see Fig.~\ref{fig:cycle-leak}). 
With this method, by adjusting the height of SH3, 
the effective leak rate to the detector could be changed. 
This method is not effected by small 
variations in shutter opening and closing times,
which can cause small systematic errors.
However, 
the effective UCN storage time 
is shorter, so the statistical sensitivity is
reduced compared to the storage method.



\section{Simulations}
\label{Sec:Simulation}

In this chapter, we present first simulations of the experiment setup, 
important for the later analysis of the data in terms of $n - n'$ oscillations. 
Several options for improvements 
of the setup used in 2020
were also simulated.

\subsection{Simulation of UCN transport and storage}
The MCUCN model~\cite{Zsigmond2018} includes the PSI UCN source and 
the UCN path up to the beamports and the experiment
with practically all details of the surfaces interacting with UCN. 
The relevant surface parameters of the UCN optics and the flux obtained 
at PSI up to the position of the beamport were calibrated with dedicated 
test measurements 
on the West-1 beamline, 
reported in Refs.~\cite{Bison2020,Ries2016,Bison2021PP}. 
Here we present an adaptation to the new $n - n'$ experiment.

\begin{figure}[htb]
\centering
\includegraphics[scale=1.20]{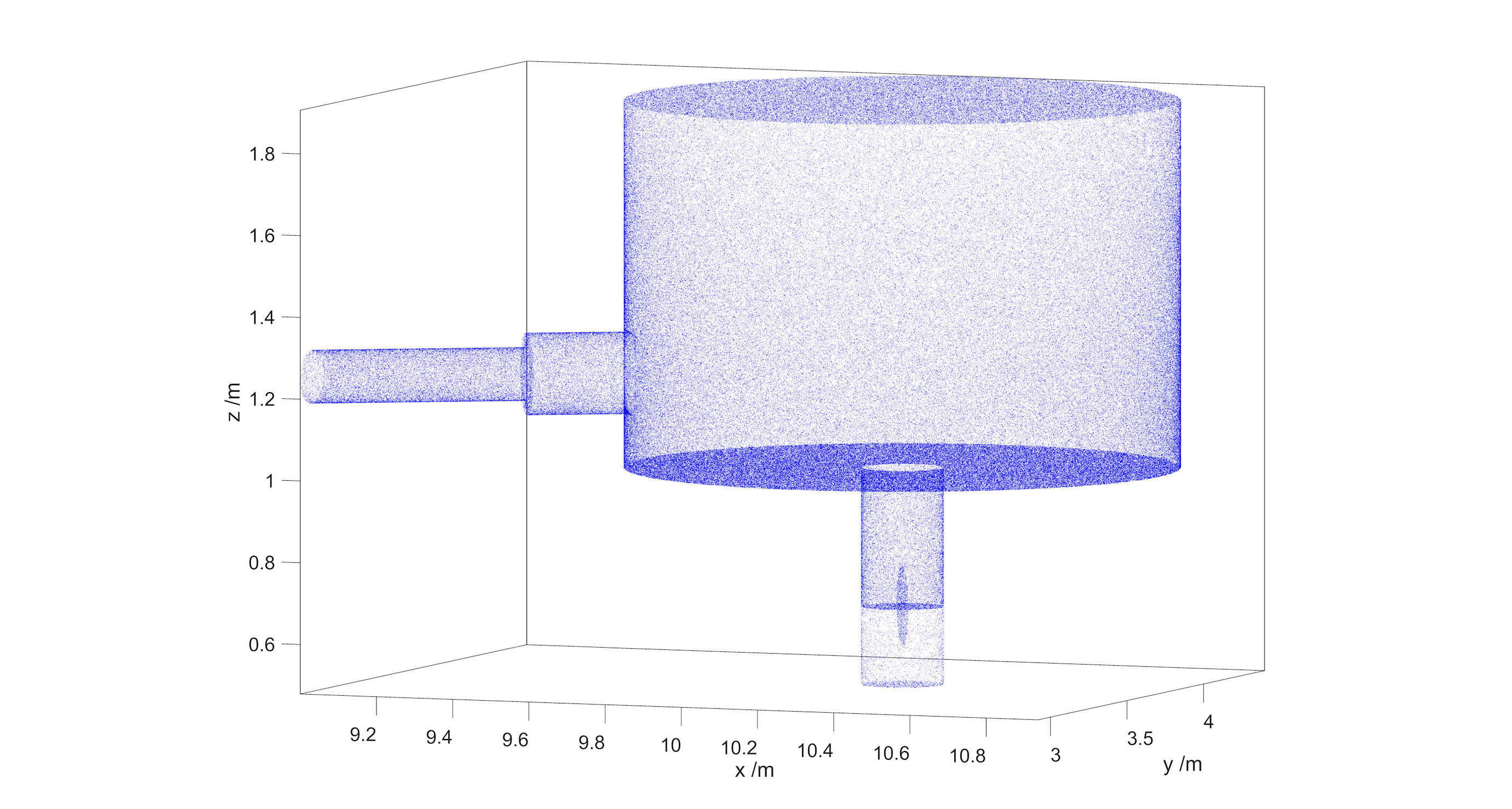}
\caption[MCUCN model]{
MCUCN model of the $n - n'$ experiment at the PSI UCN source. 
The blue dots are calculated reflection points of the UCN illuminating the guides and the storage volume.
The open shutter towards the detector is visible.
The two larger cylindrical volumes on the side and the bottom represent 
the ``pocket'' volumes as explained in the text.
 }
\label{fig:MCUCNmodel}
\end{figure}

\begin{figure}[htb]
\centering
\includegraphics[scale=0.35]
{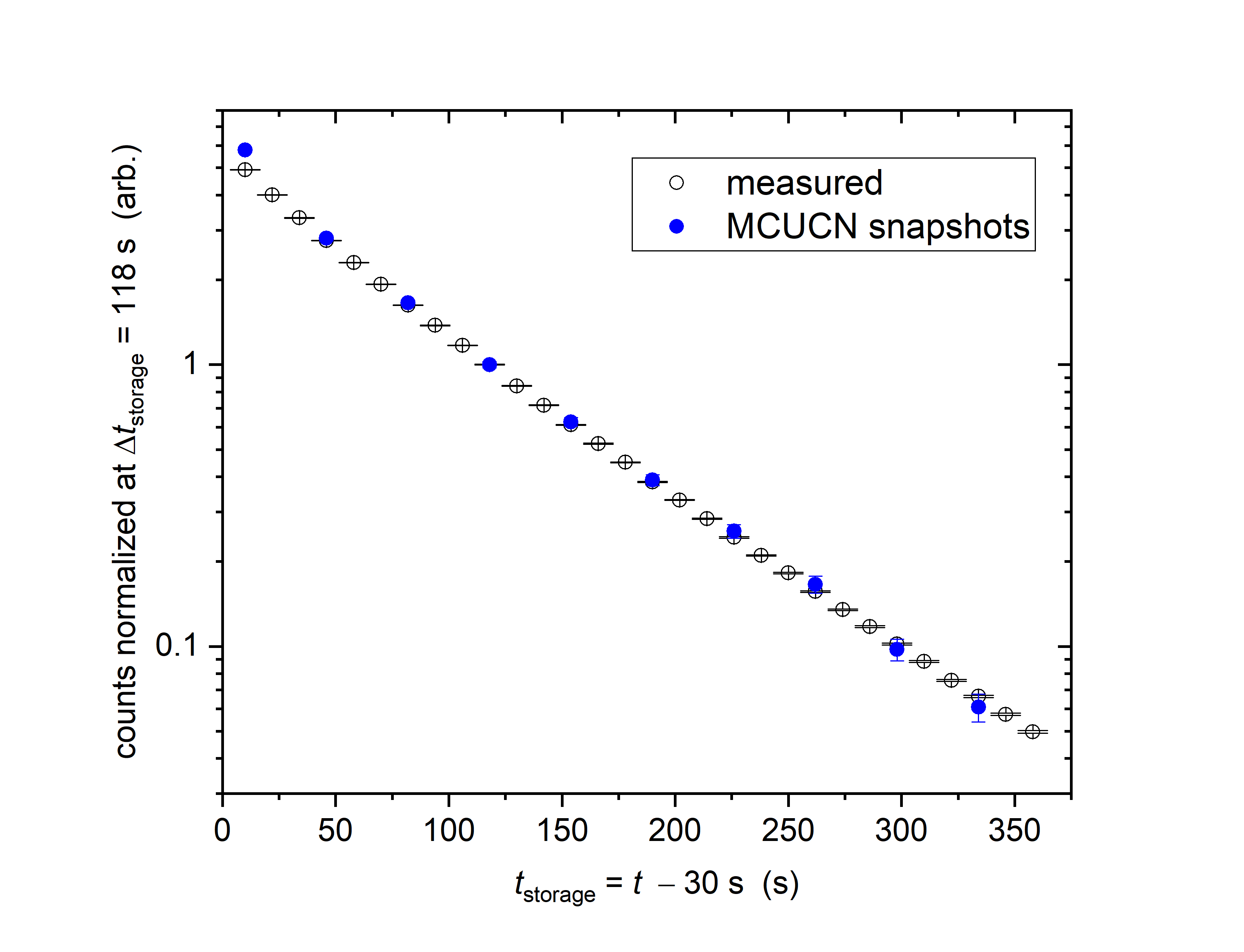}
\caption[Simulated and measured UCN storage curve]{Measured and 
simulated UCN storage curve with benchmarked parameters.
The simulation was normalized to the experimental curve for better comparison.}
\label{fig:MCUCN-StorageAfterFillingMCUCNvsMeasurement}
\end{figure}

The UCN optics model of the apparatus 
is depicted in Fig.~\ref{fig:MCUCNmodel} using calculated 
reflection points of UCNs serving as a preliminary test. 
The geometry of the simulation between the solid
deuterium surface in the UCN source and 
the beamport is not displayed here.
The experiment's geometry
consists of a 1\,m tall  
cylindrical volume
with 135.8\,cm diameter, 
adapted to connect to the beamline guide for filling. 
The VAT shutter used for filling is outside of the vacuum tank, 
creating a ``pocket'' volume during storage.
The bottom surface has an opening for the vertical guide 
leading to the detector. 
The shutter is placed in the guide below the bottom 
surface of the large chamber, constituting a second ``pocket''.
In the MC model, the guide holes can be closed instantly with the shutters. 
The detector is represented by an ideal counter below an aluminum window. 

The parameters were adjusted to match the measured storage curve shown in Fig.~\ref{fig:MCUCN-StorageAfterFillingMCUCNvsMeasurement}. We used for the experiment volume, the ``pocket'' guide parts and the detector shutter:
188\,neV neutron optical potential (stainless steel), 
loss parameter $\eta = 4 \cdot 10^{-4}$, 
(from which the loss per bounce is calculated in the 
standard way \cite{Golub1991}),
gaps around the guide entrances with gapsize $t = 3.3$~mm each
(represented by totally absorbing rings with an inner radius $r_\mathrm{guide}$, outer radius $r_\mathrm{guide} + t$),
and a 20\% fraction of Lambert diffuse reflections. The latter number is based on roughness measurements of stainless steel samples indicating about ten times larger RMS than for glass~\cite{Bison2020}, and for glass substrates Lambert reflections were benchmarked as < 2\%. 

The parameters of the VAT shutter were: 
220\,neV optical potential, 
$3 \cdot 10^{-4}$ loss parameter, 
and a 2\% fraction of diffuse reflections. 
The absorption in the detector window was set a factor of 2.2 larger than the 
loss cross section in pure aluminum as found in Ref.~\cite{Bison2020}.

\begin{figure}[htb]
\centering
\includegraphics[scale=0.35]{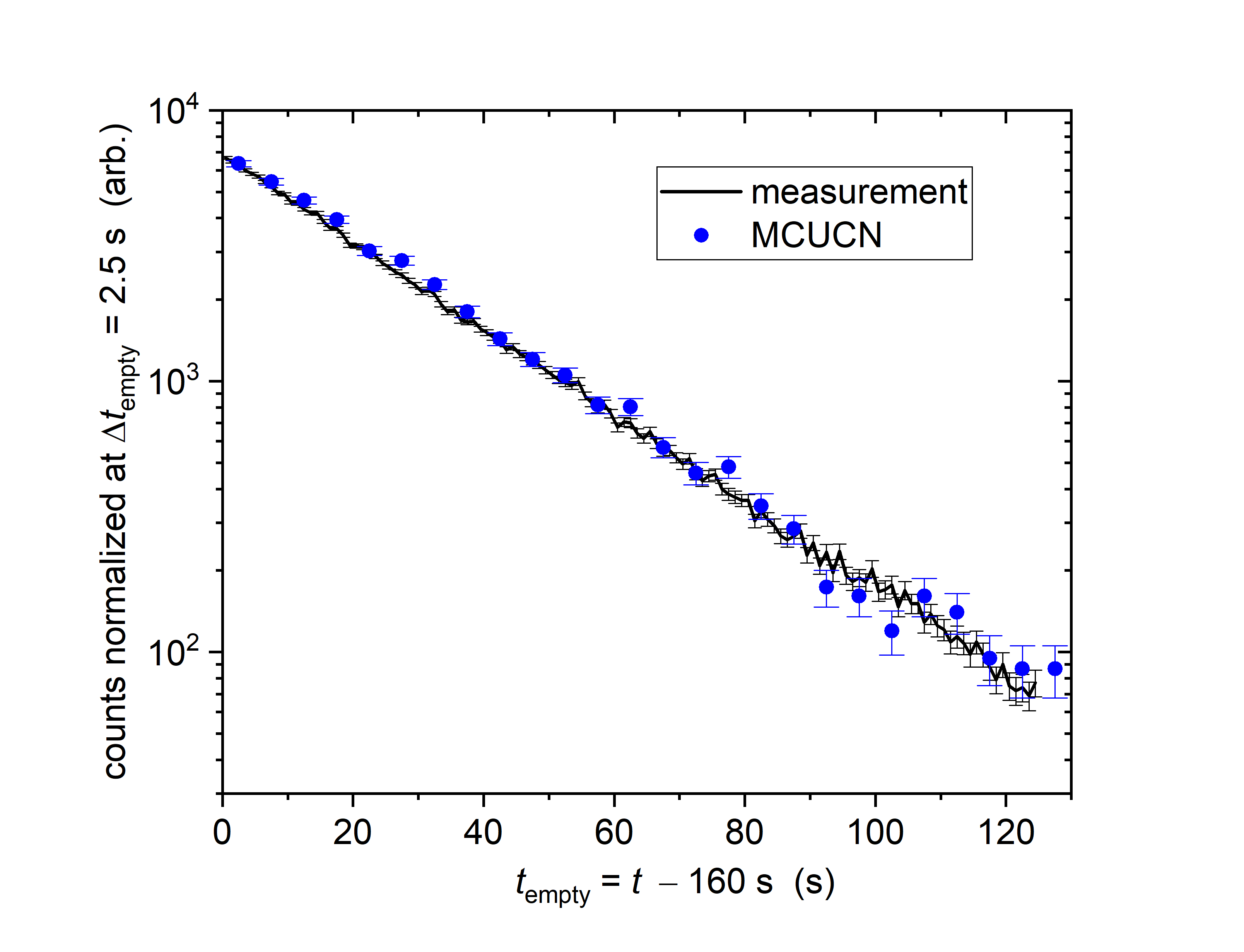}
\caption[Measured and simulated UCN emptying curves]{
Measured and simulated UCN emptying curves validating that the parameters benchmarked from the storage curves are correct.}
\label{fig:MCUCN-EmptyAfterStorageMCUCNvsMeasurement}
\end{figure}

As a control, a  simulation of the emptying curve was performed 
based on the parameters fitted to the storage curve and compared 
to the measured emptying curve 
in Fig.~\ref{fig:MCUCN-EmptyAfterStorageMCUCNvsMeasurement}.  
The good match of both profiles, using only one intensity 
scaling parameter, demonstrates the validity of the obtained parameters.

The simulations assume an 8\,s long proton beam pulse and 
a filling time of 30\,s 
optimized empirically during the test.
The emptying curve is computed for a storage time $t_s$ of 120\,s. 
The UCN emptying curves show the UCN counts as a function 
of time after opening the SH2 shutter to the detector
(compare Fig.~\ref{fig:cycle}, 'counting').

\begin{figure}[htb]
\centering
\includegraphics[scale=0.35]{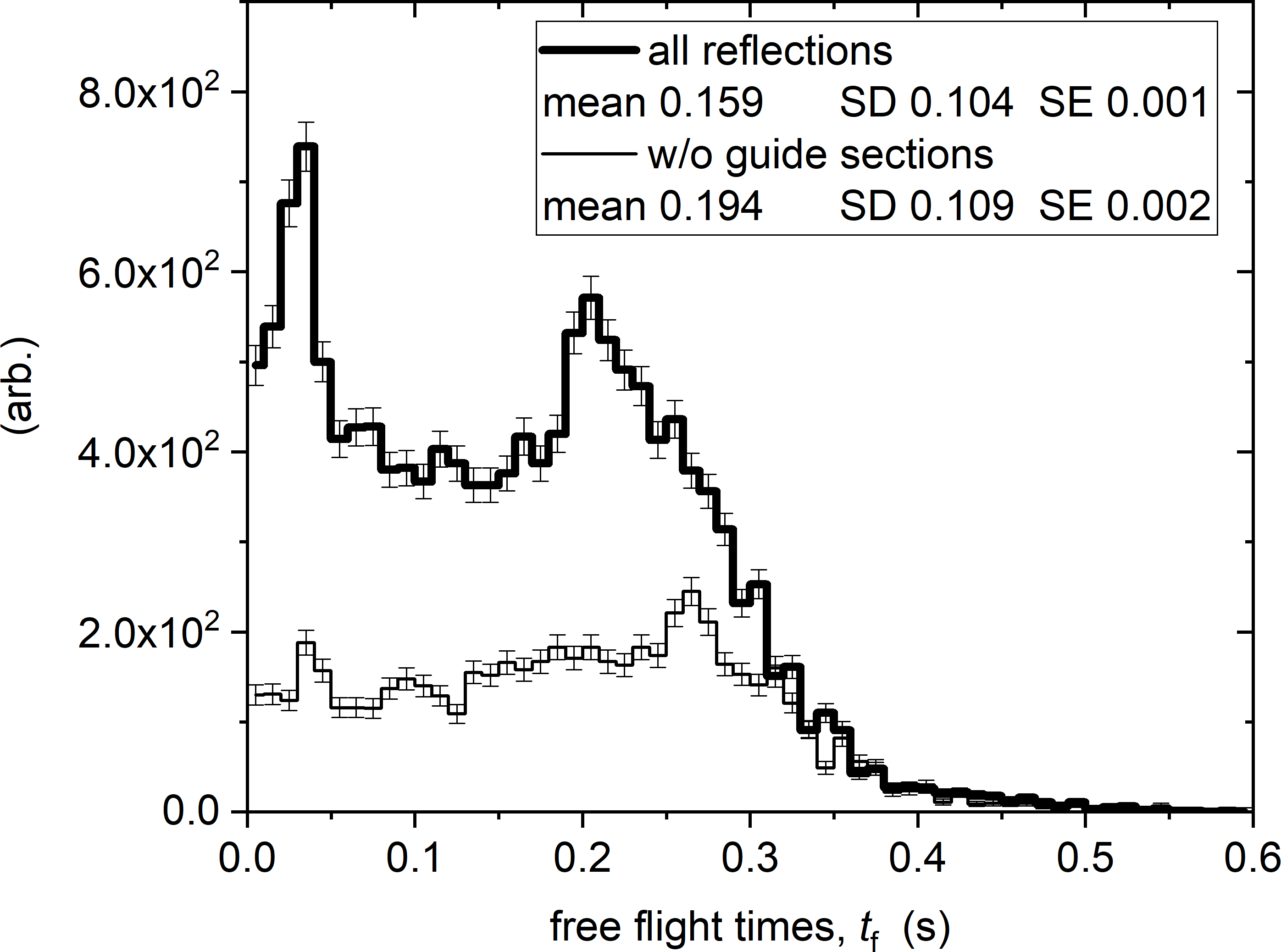}
\caption[Distribution of the free-flight-time $t_f$]{
Distribution of the free-flight-time $t_f$ 
histogrammed between 115 and 120\,s of storage. 
The thin curve represents the distribution for UCN 
that only bounce at the wall of the large chamber, i.e. without considering reflections 
in the 
attached guide sections (``pockets'').
The values indicated in the legend are:
mean - mean value of $t_f$,
SD - standard deviation, and
SE - standard error of mean.}
\label{fig:MCUCN-tf-distribution}
\end{figure}

The simulation also provided a large sample of free-flight-times, 
$t_f$, between wall collisions. 
The distribution is shown 
in Fig.~\ref{fig:MCUCN-tf-distribution} for 120\,s storage time. 
The $t_\text{f}$ samples were gathered in a 5\,s interval 
before opening the shutter to the detector, 
weighted with the UCN energy spectrum at this time.
The value $\left< t_\text{f} \right>$ is required to calculate the 
probability distribution of 
the neutron to mirror-neutron oscillation $\tau_{nn'}$ in the 
asymmetry channel
(see Eqn.~\ref{eq:asymmetry}, \ref{eq:taunnprime} and \ref{eq:Easymmetry}).

\subsection{Evaluation of upgrades to UCN components}
\label{sec:upgradesim}

Further simulations were performed in order to examine how the 
UCN statistics could be improved. 
Five different options were simulated:

\begin{enumerate}

\item Coating the shutter SH2 and tube towards the detector with NiMo $V_F$=220\,neV.

\item Installation of a new plate shutter flush with the storage volume bottom.

\item Upgrade of the filling guide to a 200\,mm large diameter
stainless steel tube from the beamport B\_SH to the entrance shutter SH1.

\item Upgrade of the filling guide to a 180\,mm NiMo-coated glass guide from 
the beamport B\_SH to the entrance shutter SH1.

\item Raising the entire setup by 0.5\,m as plotted 
in Fig.~\ref{fig:MCUCN-mirrorneutronsetup-elevated0.5m-view}
using a 200\,mm diameter stainless steel filling guide. 

\end{enumerate}

\begin{figure}[htb]
\centering
\includegraphics[scale=0.35]{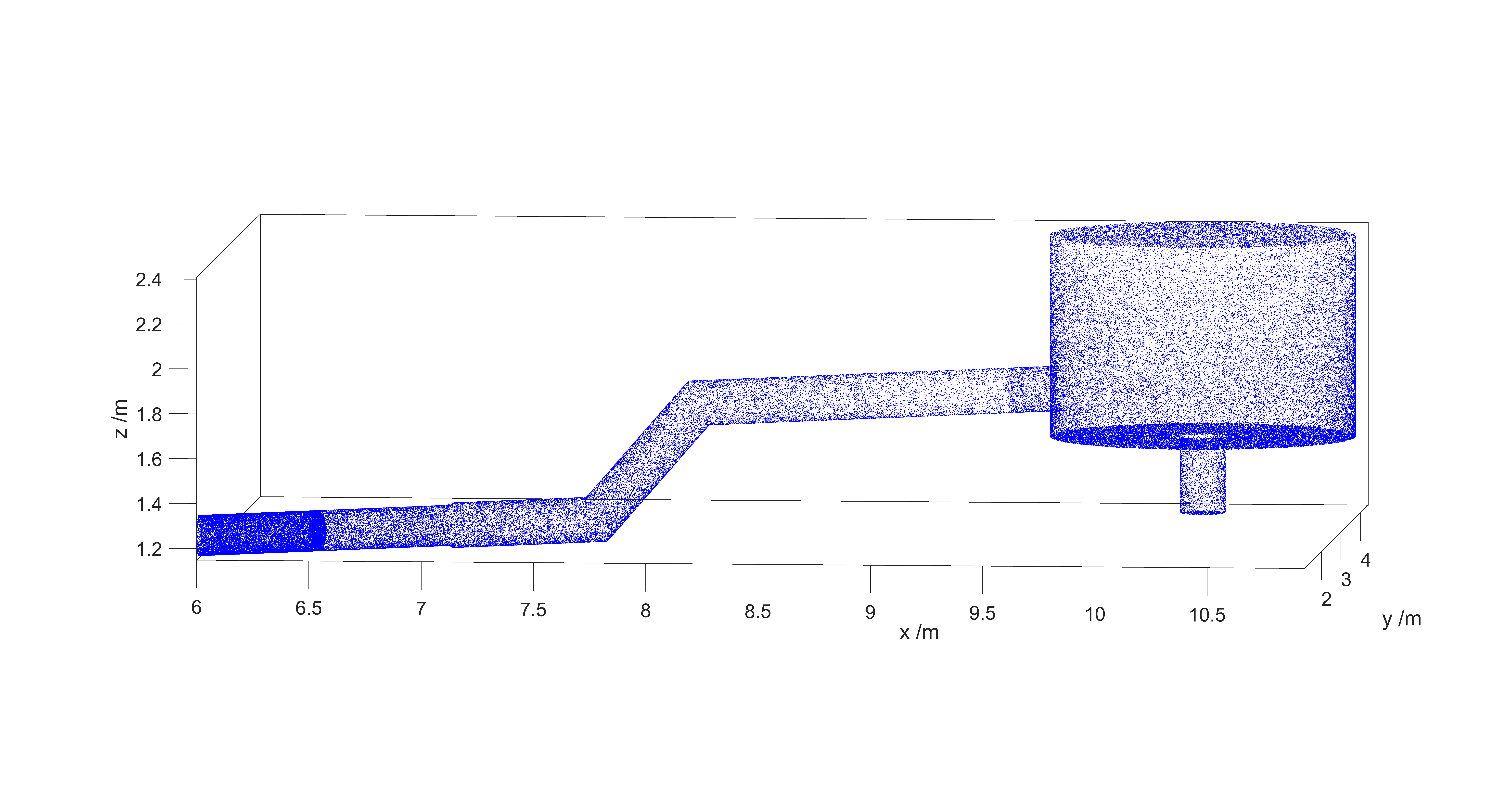}
\caption[Simulation of 0.5\,m elevated setup]{
Simulated reflection points in a MCUCN setup 
with 0.5\,m elevated UCN storage chamber using uncoated stainless steel guides.}
\label{fig:MCUCN-mirrorneutronsetup-elevated0.5m-view}
\end{figure}

The gain factors at different storage times are summarized 
in Fig.~\ref{fig:MCUCN-GainFactorsVariousCases}. 
The reference simulation is the case 
with a 130\,mm diameter NiMo coated glass UCN guide for filling
as used in the fall 2020 measurements.
Options 1, 2 and 3 can improve the number of UCNs by a factor 
$\sim$1.3 each at storage time 120\,s. 
The open symbols represent a test calculation for 
the guide walls made by NiMo on glass indicating that the losses in case 3 are 
mainly because of the large back-reflection in the rough stainless steel guides 
(in the calculation we used 20 \% Lambertian reflections). 
Option 4 gives a factor 1.5 gain with a 180\,mm NiMo on glass guide. 
Option 5 gives a factor 1.8 gain over the reference configuration.

\begin{figure}[htb]
\centering
\includegraphics[scale=0.35]{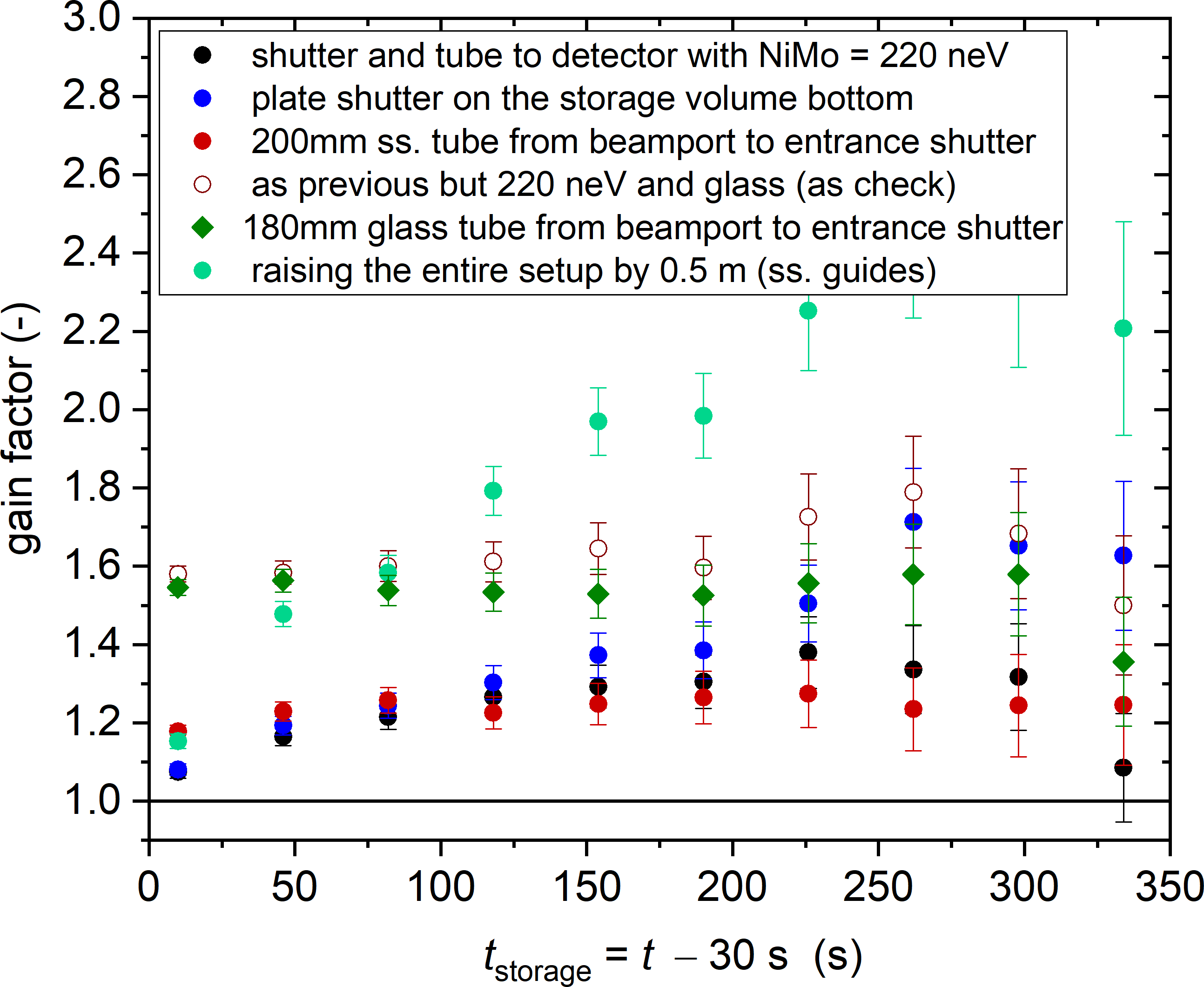}
\caption[Simulation gain factors]{
Gain factors for various improvement scenarios,
with respect to storage time,
see text for details.}
\label{fig:MCUCN-GainFactorsVariousCases}
\end{figure}

\subsection{Projected sensitivity based on 2020 measurements and field simulations}
\label{sec:projsens}

Based on the measured sensitivity of the asymmetry in counts for B-up and B-down field configurations we can estimate the 95\% C.L. for the characteristic oscillation time.
Since we have an inhomogeneous magnetic field in the chamber, 
we used in the simulation a field map obtained from previous coil calculations as illustrated in Fig.~\ref{fig:NickKickOffSlide10}.
We included a background field and assumed perfect symmetry in the B-up and B-down fields. 
The calculation method of the sensitivity in $\tau$ for the inhomogeneous 
field case was as described in 
detail in Refs.~\cite{Berezhiani:2017jkn, Biondi2018}.
With 670 cycles per $B$ field value, obtained in about 3.5 days, 
one can test the potential signals as shown 
in Fig.~\ref{fig:SimulatedLimitWest1AsymmetryNNprime}, 
for example in a 1.0 $\micro$T interval at $B'$=12.5 $\micro$T 
or a 1.5 $\micro$T interval at 25 $\micro$T.

\begin{figure}[htb]
	
	\centering

	\includegraphics[scale=0.4]{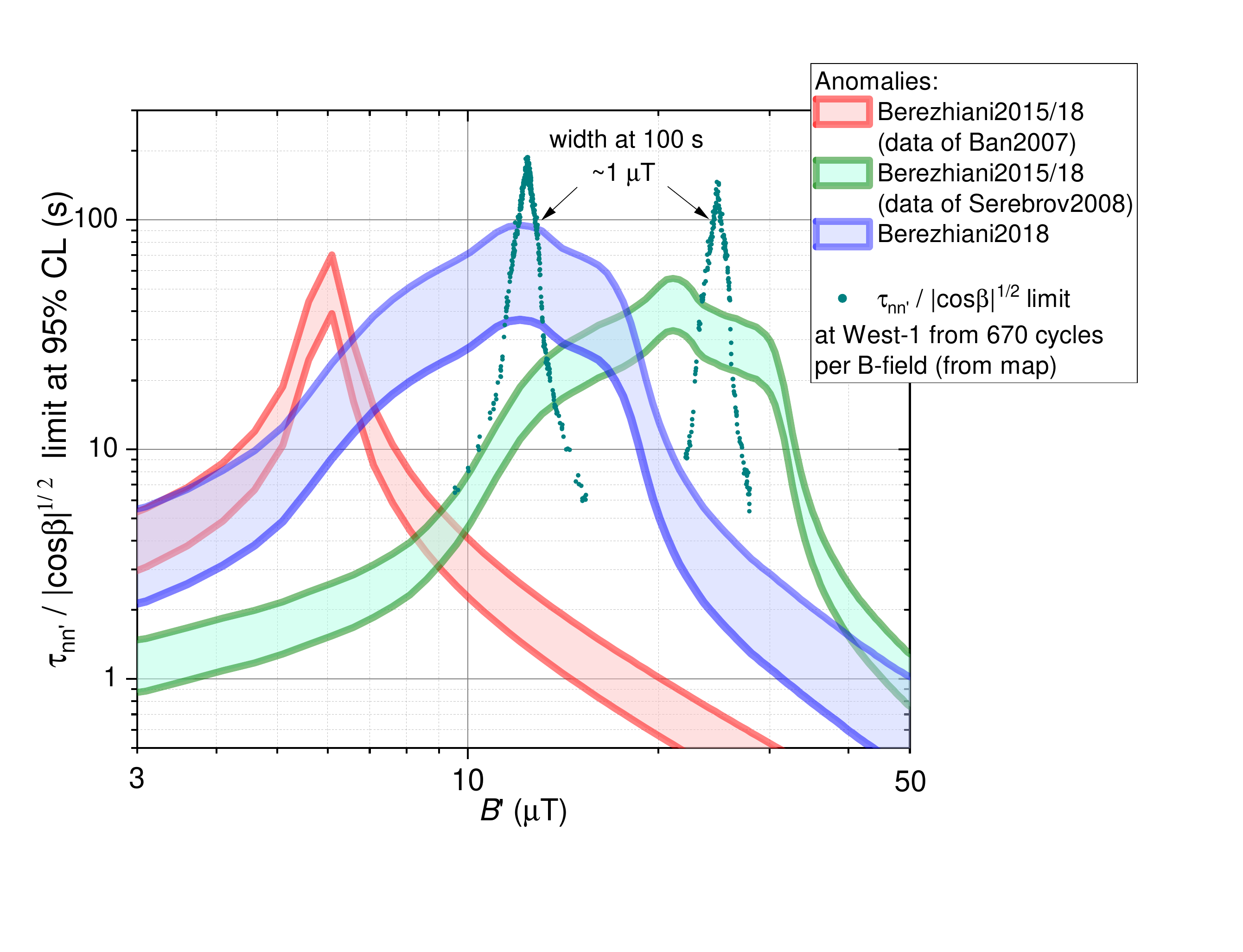}
		\caption[Simulated exclusion curves at 95\% C.L. for 670 cycles at 12.5 $\micro T$ and 25 $\micro T$.]{
		Simulated exclusion curves at 95\% C.L. corresponding to 670 cycles,
		equivalent to 3.5 days of data taking, 
		per field value for two examples of the central field, 12.5 $\micro T$ and 25 $\micro T$. 
The claimed potential signals were also plotted as 
reference represented by the bands bordered by the lines of same color.}
	
	\label{fig:SimulatedLimitWest1AsymmetryNNprime}
	
\end{figure}


\clearpage

\section{Measurement plans}
\label{Sec:BeamRequest}

The upgrade of the apparatus was carried out 
in spring 2021.
The collaboration has been taking data
for 14 weeks of beamtime
at the PSI West-1 UCN beamport in 2021.

During the main physics data-taking campaign
we focused on the regions compatible with potential signals reported 
by previous experiments, illustrated in Fig.~\ref{fig:nnp_asym_lim} 
by the red 
and gray bands. 
We used the storage technique illustrated in Fig.~\ref{subsec:Concept} to 
measure several cycles with the applied magnetic field inverted in 
a pattern to allow the compensation of all relevant drifts in the neutron counts. 
This technique was successfully demonstrated in the test measurement campaign.

The field ranges to be scanned, in order of priority, were:
\begin{enumerate}
	\item \SI{12}{\micro T} - \SI{19}{\micro T}, compatible with potential signals 
	reported by two previous experiments: \cite{Serebrov2008mirror} and \cite{Berezhiani:2017jkn}, where the red and grey 
	bands in Fig.~\ref{fig:nnp_asym_lim} overlap;
	\item \SI{22}{\micro T} - \SI{40}{\micro T} and \SI{5}{\micro T} - \SI{10}{\micro T}, compatible with the potential signals found in \cite{Serebrov2008mirror} and ~\cite{Ban:2007tp} respectively;
	\item \SI{40}{\micro T} - \SI{360}{\micro T}, where experimental bounds are very weak and could be substantially improved by a short data-taking run.
\end{enumerate}

\clearpage

%
%
%
\section{Summary}
\label{Sec:Summary}

In this paper we present a new dedicated experiment to search for 
possible neutron to mirror-neutron oscillations in the presence of mirror magnetic fields. 
A first iteration was constructed and tested in fall 2020. 
Following an upgrade program and a detailed measurement of the magnetic field 
within the UCN storage chamber, a main data-taking run is underway in 2021 
at the West-1 beamline of the PSI UCN source. 
With a 14 week measurement period, we aim to either confirm a signal at a given 
mirror magnetic field or to fully exclude a large relevant 
and as of yet not constrained parameter space 
consistent with previously reported potential signals.

\section{Acknowledgements}

The realization of the apparatus would not have been
possible without the dedicated efforts and professional support 
of Michael Meier and Luke Noorda.

Special thanks go to 
B.~Blau, P.~Erisman, S.~Gr\"unberger, 
S.~Hauri,
B.~Jehle, 
K.~Lojek,
M.~M\"ahr,  
O.~Morath, 
R.~Nicolini, 
R.~Schwarz,
B.~Zehr.

The authors acknowledge the valuable support 
of many support groups at PSI, especially
the BSQ group, the accelerator operating crews
and the `Hallendienst'.
We are grateful for support from the ETH D-PHYS vocational training division.

This work was supported by ETH Career Seed Grant SEED-13 20-2 and the 
SNF spark programme grant CRSK-2\_196416.
Dedicated funding from National Science Centre, Poland, 
grant No. 2016/23/D/ST2/00715, No. 2018/30/M/ST2/00319, and No. 2020/37/B/ST2/02349 is acknowledged.
This work is also supported by Grant G0D0421N of the Flemish Science Foundation FWO,
by the MIUR Grant under the PRIN 2017 program 2017X7X85K, 
The dark universe: synergic multimessenger approach?,
and by the SRNSF, Grant DI-18-335 "New Theoretical Models for Dark Matter Exploration".









\end{document}